\documentclass[prb,twocolumn,superscriptaddress,floatfix,longbibliography]{revtex4-1}
\usepackage[english]{babel}
\usepackage{graphicx}
\usepackage{amsmath}
\usepackage{amssymb}
\usepackage{color}
\usepackage[normalem]{ulem}
\usepackage{hyperref}
\newcommand{\im}{\mathrm{i}}
\newcommand{\dif}{\mathrm{d}}

\newcommand{\e}{\mathrm{e}}
\newcommand{\h}{\mathrm{h}}
\newcommand{\cb}{\mathrm{c}}
\newcommand{\vb}{\mathrm{v}}
\newcommand{\mrd}{\mbox{\hspace*{0.6mm}}}

\begin{document}
\title{Origin of space-separated charges in photoexcited organic heterojunctions on ultrafast time scales}
\author{Veljko Jankovi\'c} \email{veljko.jankovic@ipb.ac.rs}
\affiliation{Scientific Computing Laboratory, Center for the Study of Complex
Systems, Institute of Physics Belgrade, University of Belgrade, 
Pregrevica 118, 11080 Belgrade, Serbia}
\author{Nenad Vukmirovi\'c} \email{nenad.vukmirovic@ipb.ac.rs}
\affiliation{Scientific Computing Laboratory, Center for the Study of Complex
Systems, Institute of Physics Belgrade, University of Belgrade, 
Pregrevica 118, 11080 Belgrade, Serbia}

\begin{abstract} 
We present a detailed investigation of ultrafast (subpicosecond) exciton dynamics in the lattice model of a donor/acceptor heterojunction.
Exciton generation by means of a photoexcitation, exciton dissociation, and further charge separation are treated on equal footing.
The experimentally observed presence of space-separated charges at $\lesssim 100$ fs after the photoexcitation
is usually attributed to ultrafast transitions from excitons in the donor to charge transfer and charge separated states.
Here, we show, however, that the space-separated charges appearing on $\lesssim 100$-fs time scales are predominantly directly optically generated.
Our theoretical insights into the ultrafast pump-probe spectroscopy challenge usual interpretations of pump-probe spectra
in terms of ultrafast population transfer from donor excitons to space-separated charges. 
\end{abstract}

\maketitle{}

\section{Introduction}
\label{Sec:intro}
The past two decades have seen rapidly growing research efforts in the field of organic
photovoltaics
(OPVs), driven mainly by the promise of economically viable and environmentally
friendly power
generation.~\cite{ChemRev.110.6736,RepProgPhys.73.096401,PCCP.16.20291,
AnnRevPhysChem.66.305,PCCP.17.28451}
In spite of vigorous and interdisciplinary
research activities, there is a number
of fundamental questions that still have to be properly answered in order to
rationally design more efficient OPV devices.
It is commonly believed~\cite{ChemRev.110.6736,AcChemRes.42.1691} that photocurrent generation in OPV devices is a series of the
following sequential steps. Light absorption in the donor material
creates an exciton, which subsequently diffuses towards the donor/acceptor (D/A)
interface where it dissociates producing an interfacial charge transfer (CT) state.
The electron and hole in this state are tightly bound and localized at the D/A interface.
The CT state further separates into a free electron and a hole (the so-called charge-separated (CS) state), which are then transported to the respective electrodes.
On the other hand, several recent spectroscopic studies~\cite{nmat12-29,nmat12-66,science343-512,JAmChemSoc.136.1472}
have indicated the presence of spatially separated electrons and holes on ultrafast ($\lesssim 100$ fs) time scales after the photoexcitation.
These findings challenge the described picture of free-charge generation in OPV devices as the following issues arise.  
(i) It is not expected that an exciton created in the donor can diffuse in such a short time to the D/A interface since the distance it can cover
in 100 fs is rather small compared to the typical size of phase segregated domains in bulk heterojunctions.~\cite{afm22-1116}
(ii) The mechanism by which a CT state would transform into a CS state is not clear.
The binding energy of a CT exciton is rather large~\cite{AcChemRes.42.1691,ADMA:ADMA201000376} and there is an energy barrier
preventing it from the transition to a CS state, especially at such short time scales. 

To resolve question (ii), many experimental~\cite{nmat12-29,nmat12-66,science335-1340,jacs.135.18502}
and theoretical~\cite{FD.163.377,PhysRevB.88.205304,PhysRevB.90.115420,JPhysChemC.119.15028,PhysRevB.91.201302}
studies have challenged the implicit assumption that the lowest CT state is involved in the process.
These studies emphasized the critical role of electronically hot (energetically higher) CT states
as intermediate states before the transition to CS states.
Having significantly larger electron-hole separations, i.e., more delocalized carriers, compared to the interface-bound CT states,
these hot CT states are also more likely to exhibit ultrafast charge separation and thus bypass the relaxation to the lowest CT state.
The time scale of the described hot exciton dissociation mechanism is comparable to the time scale of hot CT exciton relaxation
to the lowest CT state.~\cite{nmat12-66,JPhysChemC.119.15028}
Other studies suggested that electron delocalization
in the acceptor may reduce the Coulomb barrier~\cite{science343-512,jacs.135.16364,PhysChemChemPhys.16.20305} and allow the transition from CT to CS states.
Experimental results of Vandewal et al.,~\cite{nmat13-63}
who studied the consequences of the direct optical excitation of the lowest
CT state, suggest that the charge separation can occur very efficiently from this state.
To resolve issue (i), it has been proposed that a direct transition from donor excitons to CS states
provides an efficient route for charge separation.~\cite{ncomms5-3119,jacs.136.2876}

All the aforementioned studies implicitly assume that an optical excitation creates a donor exciton and address the mechanisms by which it can evolve into
a CT or CS state on a $\sim 100$ fs time scale. In this work, we demonstrate that the majority of space-separated charges that are present
$\sim 100$ fs after photoexcitation are directly optically generated, in contrast to the usual belief that they originate from optical generation of donor excitons
followed by some of the proposed mechanisms of transfer to CT or CS states. We note that in a recent theoretical work Ma and Troisi~\cite{ADMA:ADMA201402294}
concluded that space-separated electron-hole pairs significantly contribute to the absorption spectrum of the heterojunction, suggesting the possibility of
their direct optical generation. A similar conclusion was also obtained in the most recent study of D'Avino et al.~\cite{jpcl.7.536}
These works, however, do not provide information about the relative importance of direct optical generation of space-separated charges in comparison to other
hypothesized mechanisms of their generation. On the other hand, in the framework of a simple, yet physically grounded model,
we simulate the time evolution of populations of various exciton states
during and after optical excitation. Working with a model Hamiltonian whose parameters have clear physical meanings,
we are able to vary model parameters and demonstrate that these variations do not violate our principal conclusion
that the space-separated charges present at $\sim 100$ fs following photoexcitation originate from direct optical generation.
In addition, we numerically investigate the ultrafast pump-probe spectroscopy and find that the signal on ultrafast
time scales is dominated by coherences rather than by state populations.
This makes the interpretation of the experimental spectra in terms of state populations rather difficult.

The paper is organized as follows. Section~\ref{Sec:theo_fram} introduces the model, its parametrization, and
the theoretical treatment of ultrafast exciton dynamics. The central conclusion of our study is presented in
Sec.~\ref{Sec:num_res}, where we also assess its robustness against variations of most of the model parameters.
Section~\ref{Sec:uf_spectro} is devoted to the theoretical approach to ultrafast pump-probe experiments and numerical
computations of the corresponding pump-probe signals. We discuss our results and draw conclusions in Sec.~\ref{Sec:discuss_conclude}.

\section{Theoretical framework}
\label{Sec:theo_fram}
In this section, we lay out the essential elements of the model (Sec.~\ref{SSec:1d_model})
and of the theoretical approach (Sec.~\ref{SSec:theo_exc_dyn}) we use to
study ultrafast exciton dynamics at a heterointerface. Section~\ref{SSec:mod_params} presents the parametrization
of the model Hamiltonian and analyzes its spectrum.
\subsection{One-dimensional lattice model of a heterojunction}
\label{SSec:1d_model}
In this study, a one-dimensional two-band lattice semiconductor model is employed to describe a heterojunction.
It takes into account electronic couplings, carrier-carrier, and carrier-phonon interactions, as well as the interaction of
carriers with the external electric field.
There are $2N$ sites in total, see Fig.~\ref{Fig:fig2}(a); first $N$ sites (labeled by $0,\dots,N-1$) belong to the donor part of the heterojunction,
while sites labeled by $N,\dots,2N-1$ belong to the acceptor part. 
Each site $i$ has one valence-band and one conduction-band orbital and also contributes localized phonon modes counted by index $\lambda_i$.
\begin{figure}[htbp]
 \begin{center}
  \includegraphics{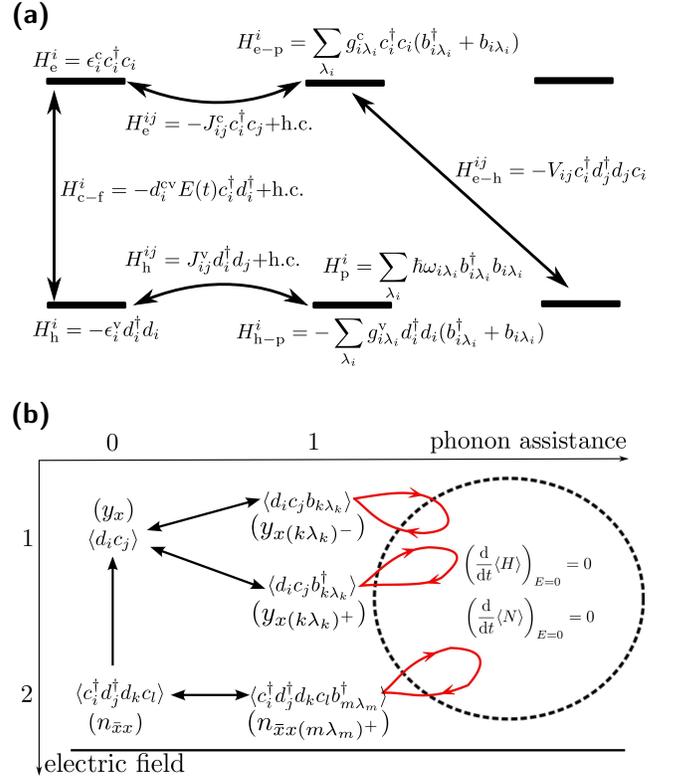}
 \end{center}
 \caption{(Color online)
 (a) Illustration of the model Hamiltonian used in our study.
 (b) Active variables in the density matrix formalism and their interrelations in the resulting hierarchy
 of equations. The direction of a straight arrow indicates that in the equation for the variable at its start appears
 the variable at its end. Loops represent couplings to higher-order phonon-assisted density matrices
 which are truncated so that the particle number and energy of the free system are conserved.}
 \label{Fig:fig1}
\end{figure}
The model Hamiltonian is pictorially presented in Fig.~\ref{Fig:fig1}(a),
the total Hamiltonian being
\begin{equation}
\label{Eq:model_H}
 H=H_\mathrm{c}+H_\mathrm{p}+H_\mathrm{c-p}+H_\mathrm{c-f}.
\end{equation}
Interacting carriers are described by
\begin{equation}
\begin{split}
 H_\mathrm{c}=\sum_{i=0}^{2N-1}\left(
H^i_\mathrm{e}+H^i_\mathrm{h}
+\sum_{\substack{j=0\\j\neq i}}^{2N-1}\left(H^{ij}_\mathrm{e}+H^{ij}_\mathrm{h}\right)
+\sum_{j=0}^{2N-1} H^{ij}_{\mathrm{e}-\mathrm{h}}\right),
\end{split}
\end{equation}
the phonon Hamiltonian is
\begin{equation}
 H_\mathrm{p}=\sum_{i=0}^{2N-1}H^i_\mathrm{p},
\end{equation}
the carrier-phonon interaction is
\begin{equation}
 H_\mathrm{c-p}=\sum_{i=0}^{2N-1}\left(H^i_{\mathrm{e}-\mathrm{p}}+H^i_{\mathrm{h}-\mathrm{p}}\right),
\end{equation}
while the interaction of carriers with the external exciting field $E(t)$ is given as
\begin{equation}
 H_\mathrm{c-f}=\sum_{i=0}^{2N-1}H^i_{\mathrm{c}-\mathrm{f}}.
\end{equation}
In Fig.~\ref{Fig:fig1}(a), Fermi operators $c_i^\dagger$ and $d_i^\dagger$ ($c_i$ and $d_i$)
create (destroy) electrons and holes on site $i$,
whereas Bose operators $b_{i\lambda_i}^\dagger$ ($b_{i\lambda_i}$) create (destroy) phonons in mode $\lambda_i$ on site $i$.
$\epsilon^\cb_i$ and $\epsilon^\vb_i$ are electron and hole on-site energies, while
$J^\cb_{ij}$ and $J^\vb_{ij}$ denote electron and hole transfer integrals, respectively.
The carrier-phonon interaction is taken to be of the Holstein form, where a charge carrier is locally and linearly coupled to
dispersionless optical modes, and $g^\cb_{i\lambda_i}$ and $g^\vb_{i\lambda_i}$ are
the interaction strengths with electrons and holes, respectively.
Electron-hole interaction is accounted for in the lowest
monopole-monopole approximation and  $V_{ij}$ is the carrier-carrier interaction potential.
Interband dipole matrix elements are denoted by $d^{\cb\vb}_i$.

\subsection{Theoretical approach to exciton dynamics}
\label{SSec:theo_exc_dyn}
We examine the ultrafast exciton dynamics during and after pulsed
photoexcitation of a heterointerface in the previously developed framework of
the density matrix theory
complemented with the dynamics controlled truncation (DCT) scheme~\cite{ZPhysB.93.195,PhysRevB.53.7244,RevModPhys.70.145}
(see Ref.~\onlinecite{PhysRevB.92.235208} and references therein),
starting from initially unexcited heterojunction.
We confine ourselves to the case of weak optical field and low carrier
densities, in which it is justified to work in the subspace
of single-exciton excitations (spanned by the so-called exciton basis) and truncate the carrier branch of
the hierarchy of equations for density matrices
retaining only contributions up to the second order in the optical field.
The phonon branch of the hierarchy
is truncated independently so as to ensure the particle-number and energy
conservation after the pulsed excitation, as described in detail in Ref.~\onlinecite{PhysRevB.92.235208}.

In more detail, the exciton basis is obtained solving the eigenvalue problem
\begin{equation}
 \label{Eq:eigen_exc_bas}
 \sum_{i'j'}\left(\delta_{i'i}\epsilon^\cb_{jj'}-\delta_{j'j}\epsilon^\vb_{ii'}
-\delta_{i'i}\delta_{j'j}V_{ij}\right)\psi^x_{i'j'}=\hbar\omega_x\psi^x_{ij},
\end{equation}
where indices $i,i'$ ($j,j'$) correspond to the position of the hole (electron) and 
quantities $\epsilon^\cb_{mn}$ ($\epsilon^\vb_{mn}$) denote on-site electron
(hole) energies (for $m=n$)
or electron (hole) transfer integrals (for $m\neq n$)
in the donor, in the acceptor, or between the donor and the acceptor.
The creation operator for the exciton in the state $x$ is then defined as
\begin{equation}
 X_x^\dagger=\sum_{ij}\psi^x_{ij} c_j^\dagger d_i^\dagger.
\end{equation}
As we pointed out,~\cite{PhysRevB.92.235208} the total Hamiltonian, in which only contributions whose
expectation values are at most of the second
order in the optical field are kept, can be expressed in terms of exciton
operators $X_x^\dagger,X_x$ as
\begin{equation}
\label{Eq:exc_ham_2nd}
\begin{split}
 H=&\sum_x \hbar\omega_x X_x^\dagger X_x+\sum_{i\lambda_i}
\hbar\omega_{i\lambda_i} b_{i\lambda_i}^\dagger b_{i\lambda_i}\\
  +&\sum_{\substack{\bar x x\\i\lambda_i}}\left(\Gamma^{i\lambda_i}_{\bar x
x}X_{\bar x}^\dagger X_x b_{i\lambda_i}^\dagger+
\Gamma^{i\lambda_i*}_{\bar x x}X_{x}^\dagger X_{\bar x} b_{i\lambda_i}\right)\\
-&E(t)\sum_x\left(M_x^* X_x+M_x X_x^\dagger\right),
\end{split}
\end{equation}
where the exciton-phonon coupling constants are given as
\begin{equation}
 \label{Eq:exc_ph_cc}
 \Gamma^{i\lambda_i}_{\bar x x}=g^\cb_{i\lambda_i}\sum_{j}\psi^{\bar
x*}_{ji}\psi^x_{ji}-g^\vb_{i\lambda_i}\sum_{j}\psi^{\bar x*}_{ij}\psi^x_{ij},
\end{equation}
while the dipole moment for the generation of the state $x$ from the ground
state is
\begin{equation}
 \label{Eq:exc_dip_mom}
 M_x=\sum_i \psi^{x*}_{ii} d^{\cb\vb}_i.
\end{equation}
Active variables in our formalism are the coherences between exciton state $x$ and the ground state, $y_x=\langle X_x\rangle$,
exciton populations (for $\bar x=x$), and exciton-exciton coherences (for $\bar x\neq x$) $n_{\bar x x}=\langle X_{\bar x}^\dagger X_x\rangle$,
together with their single-phonon-assisted counterparts
$y_{x(i\lambda_i)^-}=\langle X_x b_{i\lambda_i}\rangle$, $y_{x(i\lambda_i)^+}=\langle X_x b_{i\lambda_i}^\dagger\rangle$,
and $n_{\bar x x(i\lambda_i)^+}=\langle X_{\bar x}^\dagger X_x b_{i\lambda_i}^\dagger\rangle$.
Their mutual interrelations in the resulting hierarchy are schematically shown
in Fig.~\ref{Fig:fig1}(b), while the equations themselves  
are presented in Supplemental Material.~\footnote{See Supplemental Material at
for equations of motion of active density matrices, further results concerning the impact of model parameters on ultrafast exciton dynamics,
the mixed quantum/classical approach to exciton dynamics,
and additional details regarding numerical computations of ultrafast pump-probe spectra.}
In order to quantitatively monitor ultrafast processes at the model heterojunction
during and after its pulsed photoexcitation, the incoherent population of exciton state $x$,
which gives the number of truly bound (Coulomb-correlated) electron-hole pairs in the state $x$,
\begin{equation}
 \label{Eq:incoh_pop_x}
 \bar n_{xx}=n_{xx}-|y_x|^2,
\end{equation}
will be used. Coherent populations of exciton states, $|y_x|^2$, dominate early stages of the optical experiment, typically decay quickly due to
different scattering mechanisms (in our case, the carrier-phonon interaction), and do not represent bound electron-hole pairs.
The populations of truly bound electron-hole pairs build up on the expense of coherent exciton populations.
We frequently normalize $\bar n_{xx}$ to the total exciton population in the system,
\begin{equation}
 \label{Eq:total_ex_pop}
 N_{\mathrm{tot}}=\sum_{x}n_{xx},
\end{equation}
which, together with the expectation value of the Hamiltonian $\langle H\rangle$, is conserved in the absence of the external field.
Probabilities $f_\e(t,r)$ [$f_\h(t,r)$] that an electron (a hole) is located at site $r$ at instant $t$ can be obtained using the so-called
contraction identities (see, e.g., Ref.~\onlinecite{RevModPhys.70.145}) and are given as
\begin{equation}
 \label{Eq:def_f_e}
  f_\e(t,r)=\frac{\sum_{\bar x x}\left(\sum_{r_\h}\psi^{\bar x*}_{r_\h
r}\psi^{x}_{r_\h r}\right)n_{\bar x x}(t)}{\sum_x n_{xx}(t)},
\end{equation}
\begin{equation}
 \label{Eq:def_f_h}
  f_\h(t,r)=\frac{\sum_{\bar x x}\left(\sum_{r_\e}\psi^{\bar x*}_{r
r_\e}\psi^{x}_{r r_\e}\right)n_{\bar x x}(t)}{\sum_x n_{xx}(t)}.
\end{equation}
Consequently, the probability that an electron is in the acceptor at time $t$ is
\begin{equation}
 \label{Eq:def_prob_e_a}
 P^\e_A(t)=\sum_{r=N}^{2N-1}f_\e(t,r).
\end{equation}

\subsection{Model parameters and Hamiltonian spectrum}
\label{SSec:mod_params}
\begin{figure}[htbp]
 \begin{center}
  \includegraphics{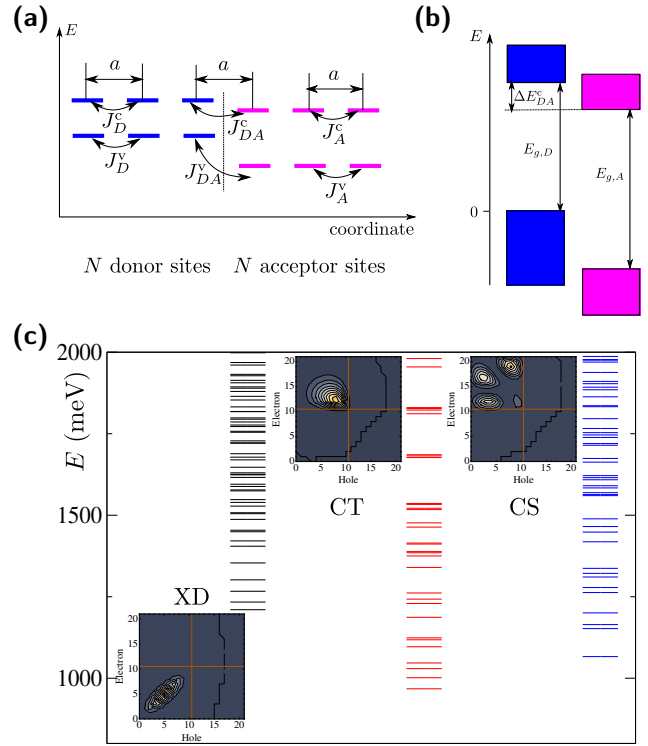}
 \end{center}
 \caption{(Color online)
 (a) One-dimensional lattice model of a heterojunction. Various types of electronic couplings
 (in the donor, in the acceptor, and among them) are indicated. There is an energy offset between single-electron/hole
 levels in the donor and acceptor. (b) Band alignment produced by our model.
 (c) Energies of exciton states, in particular of donor excitons (black lines), CT (red lines) and CS (blue lines) states.
 Exciton wave function square moduli are shown for the lowest donor, CT and CS state.}
 \label{Fig:fig2}
\end{figure}

The model Hamiltonian was parameterized to yield values of band gaps, bandwidths, band offsets, and exciton binding energies that are representative of typical OPV materials.
The values of model parameters used in numerical computations are summarized in Table~\ref{Tab:model_params}.
\begin{table}
 \caption{Values of model parameters used in calculations.}
 \label{Tab:model_params}
 \centering
 \begin{tabular}{|c c|}
  \hline
  Parameter\footnote{$E_{g,D}$ ($E_{g,A}$) is the single-particle bandgap in the donor (acceptor).
$\Delta E^\cb_{DA}$ denotes LUMO-LUMO energy offset.
$J^{\cb/\vb}_D$ ($J^{\cb/\vb}_A$) are electron/hole transfer integrals in the donor (acceptor).
$J^{\cb/\vb}_{DA}$ are electron/hole transfer integrals between the donor and acceptor.
$\varepsilon_r$ is the relative dielectric constant.
$N$ is the number of lattice sites in the donor and acceptor ($2N$ sites in total).
$a$ is the lattice constant.
$U$ denotes the on-site Coulomb interaction.
$\hbar\omega_\mathrm{p,1/2}$ are energies of local phonon modes, while
$g_{1/2}$ are carrier-phonon coupling constants.
$T$ denotes temperature. The duration of the pulse is $2t_0$.} & Value\\
  \hline
  $E_{g,D}$ (meV) & 1500\\
  $E_{g,A}$ (meV) & 1950\\
  $\Delta E^\cb_{DA}$ (meV) & 500\\
  $|J^\cb_D|$ (meV) & 105\\
  $|J^\vb_D|$ (meV) & 295\\
  $|J^\cb_A|$ (meV) & 150\\
  $|J^\vb_A|$ (meV) & 150\\
  $|J^\cb_{DA}|,|J^\vb_{DA}|$ (meV) & 75\\
  $\varepsilon_r$ & 3.0\\
  $N$ & 11\\
  $a$ (nm) & 1.0\\ 
  $U$ (meV) & 480\\
  $\hbar\omega_\mathrm{p,1}$ (meV) & 10\\
  $g_1$ (meV) & 28.5\\
  $\hbar\omega_\mathrm{p,2}$ (meV) & 185\\
  $g_2$ (meV) & 57.0\\
  $T$ (K) & 300\\
  $t_0$ (fs) & 50\\
  \hline
 \end{tabular}
\end{table}
While these values largely correspond to the PCPDTBT/PCBM interface, we note that our goal is to reach general conclusions valid for a broad class of interfaces.
Consequently, later in this study, we also vary most of the model parameters and study the effects of these variations.
Figures~\ref{Fig:fig2}(a) and~\ref{Fig:fig2}(b) illustrate the meaning of some of the model parameters.

All electron and hole transfer integrals are restricted to nearest neighbors.
The single-particle band gap of the donor $E_{g,D}$, as well as the offset $\Delta E^\cb_{DA}$ between
the lowest single-electron levels in the donor and acceptor,
assume values that are representative of the low-bandgap PCPDTBT polymer used in the
most efficient solar cells.~\cite{ADMA:ADMA200602437,ADMA:ADMA200600160}
The single-particle band gap of the acceptor $E_{g,A}$ and electron/hole transfer integrals
$J^{\cb/\vb}_{A}$ are tuned to values typical of fullerene
and its derivatives.~\cite{JPhysChemC.118.21873,PhysRevB.85.054301}
Electron/hole transfer integrals $J^{\cb/\vb}_{D}$ in the donor were extracted from the conduction and valence bandwidths of the PCPDTBT polymer.
To obtain the bandwidths, an electronic structure calculation was performed on a straight infinite polymer.
The calculation is based on the density functional theory (DFT) in the local density approximation (LDA), as implemented in the QUANTUM-ESPRESSO~\cite{QE-2009} package.
Transfer integrals were then obtained as 1/4 of the respective bandwidth. 
The values of the transfer integral between the two materials are chosen to be similar to the values obtained in the {\it ab initio} study of
P3HT/PCBM heterojunctions.~\cite{NanoLett.7.1967}
We set the number of sites in a single material to $N=11$,
which is reasonable having in mind that the typical dimensions of
phase segregated domains in bulk heterojunction morphology are considered to be $10-20$ nm.~\cite{afm22-1116} 
The electron-hole interaction potential $V_{ij}$ is modeled using the Ohno potential
\begin{equation}
 V_{ij}=\frac{U}{\sqrt{1+\left(\frac{r_{ij}}{a_0}\right)^2}},
\end{equation}
where $r_{ij}$ is the distance between sites $i$ and $j$, and $a_0=e^2/(4\pi\varepsilon_0\varepsilon_r U)$ is the characteristic length.
The relative dielectric constant $\varepsilon_r$ assumes a value typical for organic materials, while the magnitude of the on-site Coulomb interaction
$U$ was chosen so that the exciton binding energy in both the donor and the acceptor is around 300 meV.
Following common practice when studying all-organic heterojunctions,~\cite{jpcc.119.14989,bittnerramoncont}
we take one low-energy and one high-energy phonon mode. For simplicity, we assume that energies of both phonon modes, as well as
their couplings to carriers, have the same values in both materials.
The high-frequency phonon mode of energy 185 meV (approx. 1500 cm$^{-1}$), which is present in both materials, was suggested to be
crucial for ultrafast electron transfer in the P3HT/PCBM blend.~\cite{Science.344.1001}
Recent theoretical calculations of the phonon spectrum and electron-phonon coupling constants in P3HT indicate the presence of low-energy
phonon modes ($\lesssim 10$ meV) that strongly couple to carriers.~\cite{JPhysChemB.120.5572}  
The chosen values of phonon-mode energies fall in the ranges in which the phonon density of states in conjugated polymers
is large~\cite{NanoLett.9.3996} and the local electron-vibration couplings in PCBM are pronounced.~\cite{jpcc.114.20479}
We estimate the carrier-phonon
coupling constants from the value of polaron
binding energy, which can be estimated using the result of the second-order
weak-coupling perturbation theory at $T=0$ in the vicinity of the point $k=0$:~\cite{jcp.128.114713}
\begin{equation}
\label{Eq:def_e_b_pol}
 \epsilon_\mathrm{b}^\mathrm{pol}=\sum_{i=1}^2\frac{g_i^2}{2|J|}\frac{1}{\sqrt{
\left(1+\frac{\hbar\omega_{\mathrm{p},i}}{2|J|}\right)^2-1}}.
\end{equation}
We took $g_2/g_1=2$ and estimated the numerical values assuming that
$\epsilon_\mathrm{b}^\mathrm{pol}=20$ meV and $|J|=125$ meV.
The electric field is centered around $t=0$ and assumes the form
\begin{equation}
\label{Eq:wave_ramp}
 E(t)=E_0\cos(\omega_c t)\theta(t+t_0)\theta(t_0-t),
\end{equation}
where $\omega_c$ is its central frequency, $\theta(t)$ is the step function, and the duration of the pulse is $2t_0$.
The time $t_0$ should be chosen large enough so that the pulse is
spectrally narrow enough (the energy of the initially generated excitons is around the central frequency of the pulse).
On the other hand, since our focus is on processes happening on sub-picosecond time scale, the pulse
should be as short as possible in order to disentangle the carrier generation during the pulse
from free-system evolution after the pulse.
Trying to reconcile the aforementioned requirements, we choose $t_0=50$ fs.
We note that the results and conclusions to be presented do not crucially depend on the particular value
of $t_0$ nor on the wave form of the excitation. This is shown in greater detail in Supplemental Material (Supplemental Figs. 1 and 2),
where we present the dynamics for shorter pulses of wave forms given in Eqs.~\eqref{Eq:wave_ramp} and~\eqref{Eq:pump_waveform}.
Interband dipole matrix elements $d_i^\mathrm{cv}$ are zero in the acceptor ($i=N,\dots,2N-1$),
while in the donor they all assume the same value $d^\mathrm{cv}$ so that $d^\mathrm{cv}E_0=0.2$ meV (weak excitation).

Figure~\ref{Fig:fig2}(c) displays part of the exciton spectrum produced by our model.
Exciton states can be classified according to
the relative position of the electron and the hole.
The classification is straightforward only for the noninteracting heterojunction
($J^{\cb/\vb}_{DA}=0$), in which case any exciton state can be classified into
four groups:
\begin{itemize}
 \item[(a)] both the electron and the hole are in the donor [donor exciton (XD) state],
 \item[(b)] both the electron and the hole are in the acceptor (acceptor exciton state),
 \item[(c)] the electron is in the acceptor, while the hole is in the donor (space-separated exciton state),
 \item[(d)] the electron is in the donor, while the hole is in the acceptor.
\end{itemize}
Space-separated excitons can be further discriminated according to their mean
electron-hole distance defined as
\begin{equation}
 \langle r_\mathrm{e-h}\rangle_x=\sum_{ij}|i-j||\psi^x_{ij}|^2.
\end{equation}
When the electron-hole interaction is set to zero, the mean electron-hole
distance for all the states from group (c) is equal to $N$. For the non-zero
Coulomb interaction,
we consider a space-separated exciton as a CS exciton if its
mean electron-hole distance is larger than (or equal to) $N$, otherwise we
consider it as a CT exciton. In the general case, the character of an exciton state
is established by calculating its overlap with each of the aforementioned groups
of the exciton states at the noninteracting heterojunction; this state then inherits
the character of the group with which the overlap is maximal.

\section{Numerical results}
\label{Sec:num_res}
Here, the results of our numerical calculations on the model system defined in Sec.~\ref{Sec:theo_fram} are
presented.
In Sec.~\ref{SSec:int_dyn}, we observe that the populations of CT and CS states
predominantly build up during the action
of the excitation,
and that the changes in these populations occurring on $\sim$100-fs time scales after the excitation
are rather small. This conclusion, i.e., the direct optical generation as the principal source of space-separated
charges on ultrafast time scales following the excitation,
is shown in Sec.~\ref{SSec:param_vars} to be robust against variations of model parameters.
Since the focus of our study is on the ultrafast exciton dynamics at photoexcited heterojunctions,
all the computations are carried out
for 1 ps in total (involving the duration of the pulse).
\subsection{Interfacial dynamics on ultrafast time scales}
\label{SSec:int_dyn}
\begin{figure}[htbp]
 \begin{center}
  \includegraphics{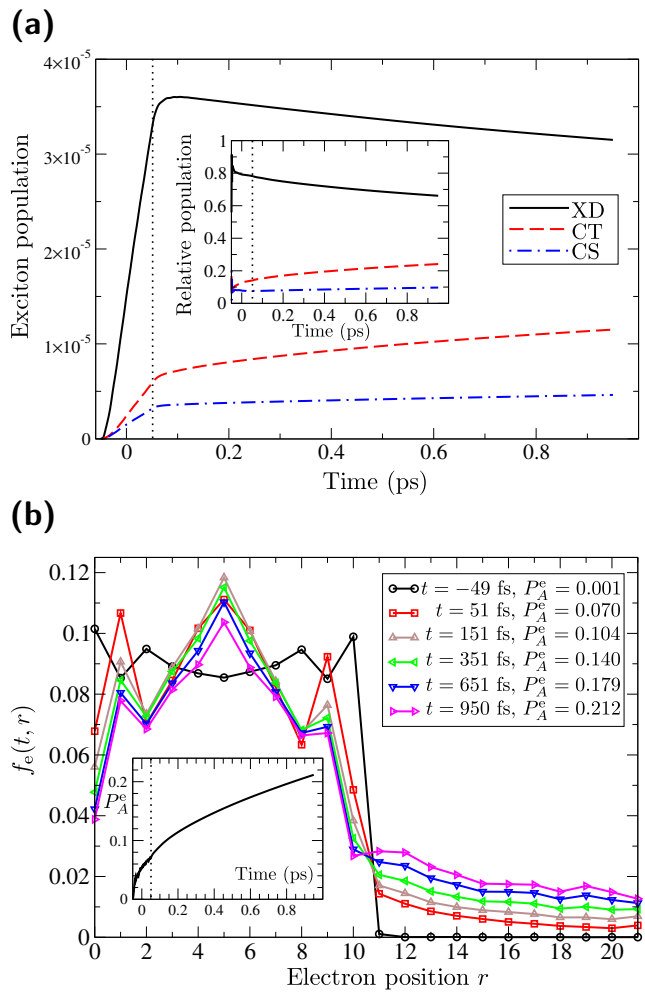}
 \end{center}
 \caption{(Color online) (a) Time dependence of the numbers of donor (XD), CT, and
CS excitons. The inset shows the time dependence of these quantities
normalized to
the total exciton population in the system.
(b) Probability that at time $t$ an electron is located at site $r$
as a function of $r$ for various values of $t$.
In the legend, the probability that at instant $t$ an electron is located in the
acceptor is given, while the inset shows its
full time dependence. Dotted vertical lines indicate the end of the excitation.}
 \label{Fig:fig3}
\end{figure}

Figure~\ref{Fig:fig3}(a) shows the time dependence of the numbers of donor,
CT, and CS excitons for the 100-fs-long excitation with
central frequency $\hbar\omega_c=1500$ meV,
which excites the system well above the lowest donor or space-separated exciton state, see Fig.~\ref{Fig:fig2}(c).
The number of all three types of excitons grows during the action of the
electric field, whereas after the electric field has vanished, the number of
donor excitons decreases and
the numbers of CT and CS excitons increase. However,
the changes in the exciton numbers brought about by the free-system evolution
alone are much less pronounced
than the corresponding changes during the action of the electric field, as is
shown in Fig.~\ref{Fig:fig3}(a). The population of CS excitons
builds up during the action of
the electric field, so that after the first 100 fs of the calculation,
CS excitons comprise 7.6\% of the total exciton population, see the inset
of Fig.~\ref{Fig:fig3}(a).
In the remaining 900 fs, when the dynamics is governed by the free Hamiltonian,
the population of CS excitons further increases to 9.6\%.
A similar, but less extreme, situation
is also observed in the relative number of CT excitons, which at
the end of the pulse form 14\% of the total population and in the remaining 900
fs of the computation their number
further grows to 24\%. 
Therefore, if only the free-system evolution were responsible for the conversion from
donor
to CT and CS excitons, the population of CT
and CS states at the end of the pulsed excitation would assume
much smaller values than we observe.
We are led to conclude that the population of CT and CS excitons on ultrafast ($\lesssim 100$-fs) time scales is mainly
established by direct optical generation.
Transitions from donor to CT and CS excitons are present, but on this time scale
are not as important as is currently thought.

Exciton dissociation and charge separation can also be monitored using the probabilities $f_\e(t,r)$
[$f_\h(t,r)$] that an electron (a hole) is located on site $r$ at instant $t$,
as well as the probability $P^\mathrm{e}_A(t)$ that an electron is in the acceptor at time $t$,
see Eqs.~\eqref{Eq:def_f_e}--\eqref{Eq:def_prob_e_a}.
Figure~\ref{Fig:fig3}(b) displays quantity $f_\e$ as a function of site index $r$ at different times $t$.
The probability of an electron being in the acceptor is a monotonically
increasing function of time $t$, see the inset of Fig.~\ref{Fig:fig3}(b).
It increases, however, more rapidly during the action of the electric field than
after the electric field has vanished: in the first 100 fs of the calculation,
it increases from virtually 0 to 0.070, while in the next 100 fs it only rises
from 0.070 to 0.104,
and at the end of the computation it assumes the value 0.210. The observed time
dependence of the probability that an electron is located in the acceptor
further corroborates our hypothesis
of direct optical generation as the main source of separated carriers on ultrafast time scales.
If only transitions from
donor to CT and CS excitons led to ultrafast charge separation starting from a donor exciton, the
values of the considered probability
would be smaller than we observe.

The rationale behind the direct optical generation of space-separated charges
is the resonant coupling between donor excitons and (higher-lying)
space-separated states, which stems from the resonant mixing between single-electron
states in the donor and acceptor modulated by the electronic coupling between materials,
see the level alignment in Fig.~\ref{Fig:fig2}(b). This mixing leads to higher-lying CT
and CS states having non-negligible amount of donor
character and acquiring nonzero dipole moment
from donor excitons; these states can thus be directly generated from the
ground state. It should be stressed that the mixing, in turn, influences donor states, which have certain amount of space-separated character.
\subsection{Impact of model parameters on ultrafast exciton dynamics}
\label{SSec:param_vars}
Our central conclusion was so far obtained using only one set of model parameters and it is therefore important to
check its sensitivity on system parameters. To this end, we vary one model parameter at a time, while all the other parameters
retain the values listed in Table~\ref{Tab:model_params}.

\begin{figure}[htbp]
 \begin{center}
  \includegraphics{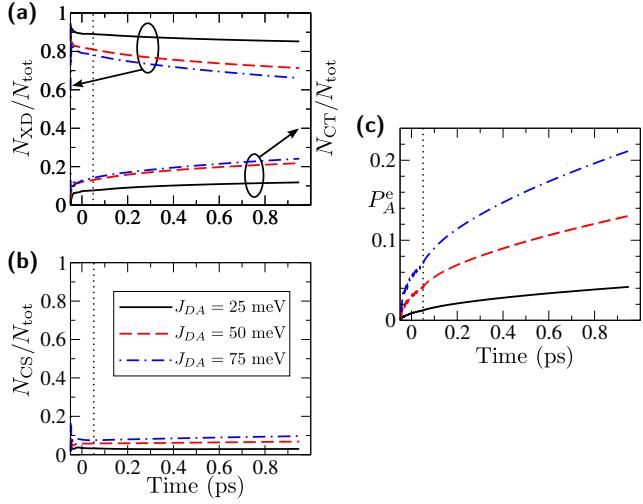}
 \end{center}
 \caption{(Color online)
Time dependence of the relative number of (a) donor and CT,
(b) CS excitons, and (c) the probability $P^\e_A$ that
an electron is in the acceptor, for
different values of the transfer integrals $|J^\cb_{DA}|=|J^\vb_{DA}|=J_{DA}$ between the
donor and the acceptor. Dotted vertical lines indicate the end of the excitation.}
 \label{Fig:fig4}
\end{figure}

We start by investigating the effect of
the transfer integral between the donor and acceptor $J^{\cb/\vb}_{DA}$.
Higher values of $J^{\cb/\vb}_{DA}$ favor charge separation, since the relative numbers of CT and CS
excitons, together with the probability that an electron is in the acceptor, increase,
whereas the relative number of donor excitons decreases with increasing $J^{\cb/\vb}_{DA}$,
see Figs.~\ref{Fig:fig4}(a)--\ref{Fig:fig4}(c).
In light of the proposed mechanism of ultrafast direct optical generation of space-separated charges, the observed trends
can be easily rationalized. Stronger electronic coupling between materials leads to stronger mixing between donor and space-separated states, i.e.,
a more pronounced donor character of CT and CS states and consequently a larger dipole moment for direct creation of CT and CS states from the ground state.

\begin{figure}[htbp]
 \begin{center}
  \includegraphics[width=0.45\textwidth]{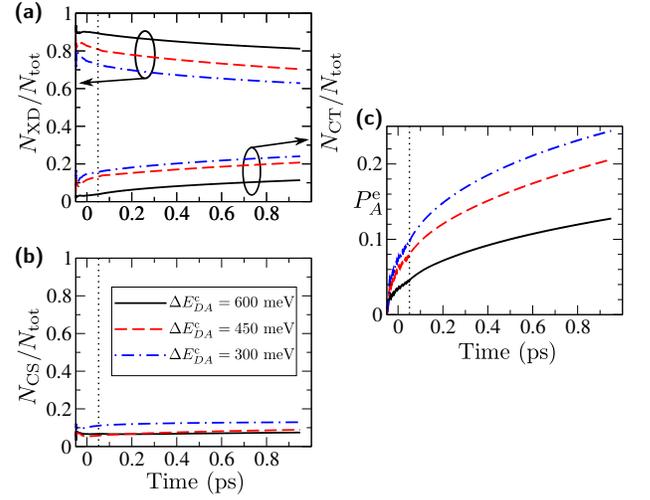}
 \end{center}
 \caption{(Color online) Time dependence of the relative number of (a) donor and CT,
(b) CS excitons, and (c) the probability $P^\e_A$ that
an electron is in the acceptor, for different values of the
LUMO-LUMO energy offset $\Delta E^\cb_{DA}$. Dotted vertical lines indicate the end of the excitation.}
 \label{Fig:fig6}
\end{figure}

The results concerning the effects of the energy offset $\Delta E^\cb_{DA}$ between LUMO levels in the donor and acceptor
are summarized in Figs.~\ref{Fig:fig6}(a)--\ref{Fig:fig6}(c).
The parameter $\Delta E^\cb_{DA}$
determines the energy width of the overlap region between single-electron states
in the donor and acceptor, see Fig.~\ref{Fig:fig2}(b). The smaller is $\Delta E^\cb_{DA}$, the
greater is the number of
virtually resonant single-electron states in the donor and in the acceptor and
therefore the greater is the number of (higher-lying) CT and CS states that
inherit nonzero dipole moments from donor states and may thus be directly
excited from the ground state. This manifests as a larger number of CT and CS excitons, as well as a larger probability
that an electron is in the acceptor, with decreasing $\Delta E^\cb_{DA}$.

\begin{figure}[htbp]
 \begin{center}
  \includegraphics[width=0.45\textwidth]{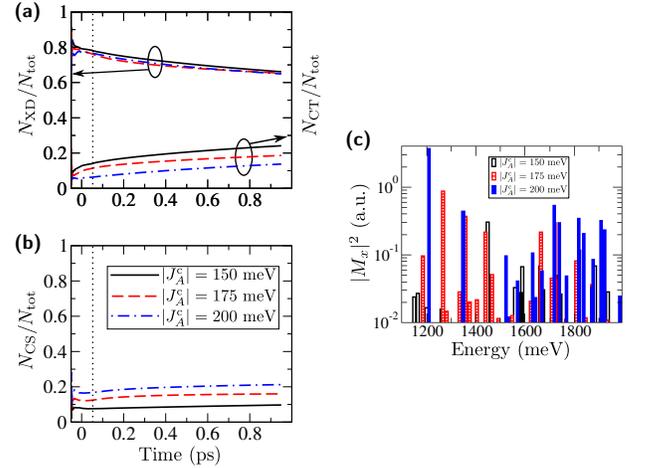}
 \end{center}
 \caption{(Color online)
Time dependence of the relative number of (a) donor and CT, and (b) CS excitons,
for different values of electronic coupling in the acceptor $J^\cb_{A}$.
(c) Squared moduli of dipole matrix elements (in arbitrary units) for direct generation of CS excitons from the ground state
for different values of electronic coupling in the acceptor $J^\cb_{A}$. Dotted vertical lines indicate the end of the excitation.
Note that, globally, squared moduli of dipole matrix elements are largest for $|J^\cb_A|=200$ meV (completely filled bars).}
 \label{Fig:fig5}
\end{figure}

Figures~\ref{Fig:fig5}(a)--\ref{Fig:fig5}(c) show the effects of electron delocalization in the acceptor on the ultrafast dynamics at the model heterojunction.
Delocalization effects are mimicked by varying the electronic coupling in the acceptor.
While increasing $|J^\cb_{A}|$ has virtually no effect on the relative number of
donor excitons, it leads to an increased participation of CS
and a decreased participation of CT excitons in the total exciton
population. CT states, in which the electron-hole interaction is rather strong, are mainly formed from
lower-energy single-electron states in the acceptor and higher-energy single-hole states in the donor. These single-particle states are not subject to
strong resonant mixing with single-particle states of the other material.
However, CS states are predominantly composed of lower-energy single-hole donor states and higher-energy single-electron acceptor states; the mixing of the latter group
of states with single-electron donor states is stronger for larger $|J^\mathrm{c}_A|$, just as in case of smaller $\Delta E^\cb_{DA}$, see Fig.~\ref{Fig:fig2}(b).
Therefore the dipole moments for direct generation of CS excitons generally increase when increasing $|J^\mathrm{c}_A|$, see Fig.~\ref{Fig:fig5}(c), whereas the dipole moments
for direct generation of CT excitons at the same time change only slightly, which can account for the trends of the participation of CS and CT excitons in
Figs.~\ref{Fig:fig5}(a) and~\ref{Fig:fig5}(b).

We now turn our attention to the effects that the strength of the carrier-phonon interaction has on the
ultrafast exciton dynamics at heterointerfaces.
\begin{figure}[htbp]
 \begin{center}
  \includegraphics{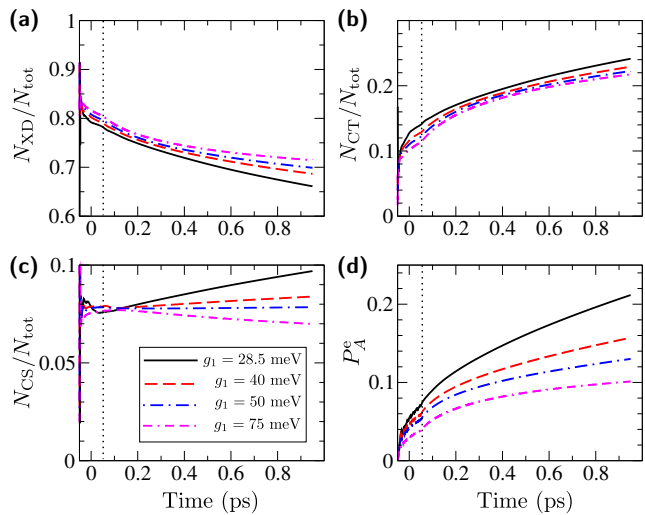}
 \end{center}
 \caption{(Color online)
Time dependence of the relative number of (a) donor, (b) CT , (c) CS excitons, and
(d) the probability $P^\e_A$ that
an electron is in the acceptor, for different strengths of the carrier-phonon interaction.
Dotted vertical lines indicate the end of the excitation.}
 \label{Fig:fig_eph}
\end{figure}
In Figs.~\ref{Fig:fig_eph}(a)--\ref{Fig:fig_eph}(d), we present the results with the fixed ratio $g_2/g_1=2.0$ and
the polaron binding energies defined in Eq.~\eqref{Eq:def_e_b_pol} assuming the values of approximately
20, 40, 60, and 140 meV, in ascending order of $g_1$.
We note that it is not straightforward to predict the effect of the variations of carrier-phonon interaction strength on the
population of space-separated states.
Single-phonon-assisted processes preferentially couple exciton states of the same character, i.e., a donor exciton state
is more strongly coupled to another donor state, than to a space-separated state.
On the one hand, stronger carrier-phonon interaction implies more pronounced exciton dissociation and charge separation
because of stronger coupling between donor and space-separated states.
On the other hand, stronger carrier-phonon interaction leads to faster relaxation of initially generated donor excitons
within the donor exciton manifold to low-lying donor states. Low-lying donor states are essentially uncoupled from
space-separated states, i.e., they exhibit low probabilities of exciton dissociation and charge separation.
Our results, shown in Figs.~\ref{Fig:fig_eph}(a)--\ref{Fig:fig_eph}(d), indicate that
stronger carrier-phonon interaction leads to smaller number of CT and CS excitons, as well as the probability
that an electron is in the acceptor, and to greater number of donor excitons.
We also note that stronger carrier-phonon interaction changes the trend displayed by the population of CS states.
While for the weakest interaction studied CS population grows after the excitation, for the strongest interaction studied
CS population decays after the excitation. This is a consequence of more pronounced phonon-assisted processes
leading to population of low-energy CT states once a donor exciton performs a transition to a space-separated state. 
This discussion can rationalize the changes in relevant quantities summarized in Figs.~\ref{Fig:fig_eph}(a)-~\ref{Fig:fig_eph}(d);
the magnitudes of the changes observed are, however, rather small.
In previous studies,~\cite{jpcc.119.14989,PhysRevB.91.041107} which did not deal with the initial
exciton generation step, stronger carrier-phonon interaction is found to suppress quite strongly the charge separation process.
The weak influence of the carrier-phonon interaction strength on ultrafast heterojunction dynamics that we observe supports the mechanism of ultrafast direct optical
generation of space-separated charges. If the charge separation process at heterointerfaces were mainly driven by the free-system evolution,
greater changes in the quantities describing charge separation efficiency would be expected with varying carrier-phonon interaction strength.

Additionally, we have performed computations for a fixed value of $\epsilon^\mathrm{pol}_\mathrm{b}$ [Eq.~\eqref{Eq:def_e_b_pol}]
and different values of the ratio $g_2/g_1$ among coupling constants of high- and low-frequency phonon modes.
The result, which is presented in Supplemental Material
(Supplemental Fig. 4),
shows that the increase of the ratio $g_2/g_1$
increases the number of CT excitons and decreases the number of donor excitons, while the population of CS states
exhibits only a weak increase. Stronger coupling to the high-frequency phonon mode (with respect to the low-frequency one)
enhances charge separation by decreasing the number of donor excitons, but at the same time promotes
phonon-assisted processes towards more strongly bound CT states, so that the population of CS states remains nearly constant.

Our formalism takes into account the influence of phonons on excitons. However, if this influence were too strong, the hierarchy of equations would have to be
truncated at a higher level, which would make it computationally intractable. When the effects of lattice motion on excitons are strong, one has, in turn,
to consider the feedback of excitons on phonons, which is not captured by the current approach. The feedback of excitons on the lattice motion can be easily
included in a mixed quantum/classical approach, where excitons are treated quantum mechanically, while the lattice motion is treated classically.
To estimate the importance of the feedback of excitons on the lattice motion, we have performed the computation using
the surface hopping approach~\cite{JChemPhys.93.1061,JPhysChemLett.5.713} (see Supplemental Material for more details).
In Supplemental Fig. 3 we show the time dependence of the probability that an electron is in the acceptor obtained from simulations with and without feedback effects.
The result is nearly the same in both cases, suggesting that feedback effects are small. As a consequence, our approach is sufficient for properly taking into
account the influence of phonons on excitons.

We have also studied the influence of the temperature on the ultrafast exciton dynamics at a heterojunction.
It exhibits a weak temperature dependence,
see Supplemental Fig. 5,
which is consistent with existing
theoretical~\cite{JChemPhys.140.044104} and experimental~\cite{JPhysChemLett.1.2255} insights, and also with the mechanism
of direct optical generation of space-separated carriers.

\begin{figure}[htbp]
 \begin{center}
  \includegraphics{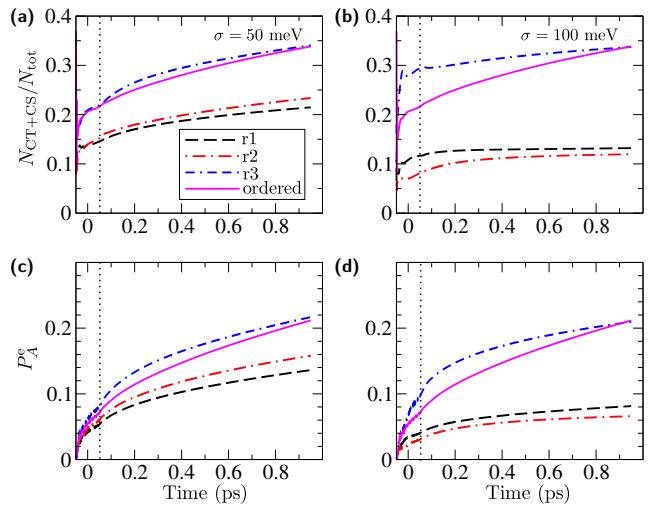}
 \end{center}
 \caption{(Color online) (a) and (b) Time dependence of the relative number of space-separated (CT and CS) excitons for three different disorder realizations (r1,r2,and r3).
 (c) and (d) Time dependence of the probability that an electron is in the acceptor for three different disorder realizations.
 Time evolution of the respective quantities in ordered system is shown on each graph for comparison.
 Disorder is diagonal (affects only on-site energies), static, and Gaussian, standard deviations being $\sigma=50$ [(a) and (c)]
and $100$ meV [(b) and (d)].}
 \label{fig:diff_disorder}
\end{figure}

Finally, the consequences of introducing diagonal static disorder in our model will be studied.
It is done by drawing the (uncorrelated) on-site energies of electrons and holes in the donor and the acceptor
from Gaussian distributions centered at the values that can be obtained from Table~\ref{Tab:model_params}.
We have for simplicity assumed that the standard deviations of all the Gaussian distributions are
equal to $\sigma$. As we do not intend to obtain any of the system properties by a statistical analysis
of various realizations of disorder, but merely to check whether or not
the presence of disorder may significantly alter qualitative features of
the proposed picture of ultrafast exciton dynamics at heterointerfaces,
we present our results only for a couple of different disorder realizations and compare them to the results for ordered system.
In Figs.~\ref{fig:diff_disorder}(a)--\ref{fig:diff_disorder}(d) we show the time dependence
of the relative number of space-separated (CS and CT) excitons and of the probability $P^\e_A$
for three different realizations of disorder with standard deviations
$\sigma=50$ and $100$ meV.
For these disorder realizations, the quantities we use to describe ultrafast heterojunction dynamics show qualitatively
similar behavior to the case of the ordered system.
Namely, changes in the relative number of space-separated excitons and the probability of an electron being in the acceptor
are more pronounced during the action of the pulse than after its end.
The characteristic time scales of these changes (for the disorder realizations studied) are not drastically different from
the corresponding time scale in the ordered system.
The presence of disorder in our model does not necessarily lead to less efficient charge separation as monitored by the
two aforementioned quantities.
Our results based on the considered disorder realizations are in agreement with the more detailed study of the effects of disorder on charge separation
at model D/A interfaces,~\cite{PhysRevB.88.205304} from which
emerged that regardless of the degree of disorder, the essential physics of free hole and electron generation remains the same.

In summary, we find that regardless of the particular values of varied model parameters
($J^{\cb/\vb}_{DA},\Delta E^\cb_{DA},J^\cb_A$, carrier-phonon coupling constants),
the majority of CT and CS states that are present at $\sim 100$ fs after photoexcitation have been directly generated during the excitation.
Trends in quantities describing ultrafast heterojunction dynamics
that we observe varying model parameters can be explained by taking into consideration
the proposed mechanism of ultrafast direct optical generation of space-separated charges.   

\section{Ultrafast spectroscopy signatures}
\label{Sec:uf_spectro}
Exciton dynamics on ultrafast time scales is typically probed experimentally using the ultrafast pump-probe spectroscopy,
see, e.g., Refs.~\onlinecite{nmat12-29,nmat12-66}.
In such experiments, the presence of space-separated charges on ultrafast time scales after photoexcitation has
been established and the energy resonance between donor exciton and space-separated states was identified as responsible for
efficient charge generation,~\cite{nmat12-29} in agreement with our numerical results.
However, while our results indicate that the majority of space-separated charges that are present at $\sim 100$ fs after photoexcitation
have been directly optically generated, interpretation of experiments~\cite{nmat12-29} suggests that these states become populated by the transition
from donor exciton states. To understand the origin of this apparent difference, we numerically compute ultrafast pump-probe
signals in the framework of our heterojunction model. In Sec.~\ref{SSec:theo_spectro}, we present the theoretical treatment
of ultrafast pump-probe experiments adapted for the system at hand.
Assuming that the probe pulse is deltalike,
we obtain an analytic expression
relating the differential transmission $\Delta T$ to the nonequilibrium state of the system ``seen" by the probe pulse.
The expression provides a very clear and direct interpretation of the results of ultrafast pump-probe experiments
and allows to distinguish between contributions stemming from exciton populations and coherences, challenging
the existing interpretations.
It is used in Sec.~\ref{SSec:num_spectro} to numerically compute differential transmission signals.
\subsection{Theoretical treatment of the ultrafast pump-probe spectroscopy}
\label{SSec:theo_spectro}
In a pump-probe experiment, the sample is firstly irradiated by an energetic
pump pulse and
the resulting excited (nonequilibrium) state of the sample is consequently
examined using a second, weaker,
probe pulse, whose time delay with respect to the pump pulse can be
tuned.~\cite{MukamelBook,AdvMater.23.5468,PhysRevA.91.033416}
Our theoretical approach to a pump-probe experiment considers the interaction with the pump pulse
as desribed in Sec.~\ref{SSec:theo_exc_dyn} and Ref.~\onlinecite{PhysRevB.92.235208}, i.e.,
within the density matrix formalism employing the DCT scheme up to the second order in the pump field.
The interaction with the probe pulse is assumed not to change significantly the nonequilibrium
state created by the pump pulse and is treated in the linear response regime. The corresponding nonequilibrium
dipole-dipole retarded correlation function is then used to calculate pump-probe signals.~\cite{PhysRevA.91.033416,EurPhysJB.89.128}
 
To study pump-probe experiments, we extended our two-band lattice semiconductor model including more
single-electron (single-hole) energy levels per site. Multiple single-electron
(single-hole) levels on each site should be dipole-coupled among themselves in order to enable
probe-induced dipole transitions between various exciton states.
We denote by $c_{i\beta_i}^\dagger$ ($c_{i\beta_i}$) creation (annihilation) operators
for electrons on site $i$ in conduction-band orbital $\beta_i$; similarly, $d_{i\alpha_i}^\dagger$ ($d_{i\alpha_i}$)
create (annihilate) a hole on site $i$ in valence-band orbital $\alpha_i$.
The dipole-moment operator in terms of electron and hole operators assumes the form
\begin{equation}
\label{Eq:P_c_d}
\begin{split}
 P&=\sum_{\substack{i\\ \beta_i\alpha_i}}\left(d^\mathrm{cv}_{i} c_{i\beta_i}^\dagger d_{i\alpha_i}^\dagger+\mathrm{h.c.}\right)\\
&+\sum_{\substack{i\\ \beta_i\neq\beta_i'}} d^\mathrm{cc}_i c_{i\beta_i}^\dagger c_{i\beta_i'}
-\sum_{\substack{i\\ \alpha_i\neq\alpha_i'}} d^\mathrm{vv}_i d_{i\alpha_i'}^\dagger d_{i\alpha_i}.
\end{split}
\end{equation}
Intraband dipole matrix elements $d^\mathrm{cc}_i$ ($d^\mathrm{vv}_i$)
describe electron (hole) transitions between different single-electron (single-hole)
states on site $i$, as opposed to the interband matrix elements $d^{\cb\vb}_i$, which are responsible for the exciton generation.
Performing transition to the exciton basis, which is defined analogously to Eq.~\eqref{Eq:eigen_exc_bas},
dipole matrix elements for transitions from the ground state to exciton state $x$ are
\begin{equation}
 M_x=\sum_{\substack{i\\ \beta_i\alpha_i}}d^\mathrm{cv}_{i} \psi_{(i\alpha_i)(i\beta_i)}^{x*},
\end{equation}
while those for transitions from exciton state $x$ to exciton state $\bar x$ are
\begin{equation}
\begin{split}
 M_{\bar x}^x&=\sum_{\substack{i\\ \alpha_i\neq\alpha_i'}}\sum_{\substack{j\\ \beta_j}}\psi_{(i\alpha_i)(j\beta_j)}^{\bar x*} d^\mathrm{vv}_{i}\psi_{(i\alpha_i')(j\beta_j)}^{x}\\
&-\sum_{\substack{i\\ \beta_i\neq\beta_i'}}\sum_{\substack{j\\ \alpha_j}}\psi_{(j\alpha_j)(i\beta_i')}^{\bar x*} d^\mathrm{cc}_i \psi_{(j\alpha_j)(i\beta_i)}^x.
\end{split}
\end{equation}
Operator $P$ [Eq.~\eqref{Eq:P_c_d}] expressed in terms of operators $X_x,X_x^\dagger$ assumes the form
(keeping only contributions whose expectation values are at most of the second order in the pump field)
\begin{equation}
\label{Eq:dip_mom_op}
 P=\sum_x\left(M_x X_x^\dagger+M_x^* X_x\right)-\sum_{\bar x x}M_{\bar x}^x X_{\bar x}^\dagger X_x.
\end{equation}

We concentrate on the so-called nonoverlapping regime,~\cite{PhysRevA.91.033416} in which the probe pulse,
described by its electric field $e(t)$, acts after the pump pulse.
We take that our system meets the condition of optical thinness, i.e., the electromagnetic field
originating from probe-induced dipole moment can be neglected compared to the electromagnetic field
of the probe. In the following considerations, the origin of time axis $t=0$ is taken to be the instant at which
the probe pulse starts. The pump pulse finishes at $t=-\tau$, where $\tau$ is the time delay between (the end of)
the pump and (the start of) the probe.
The pump creates a nonequilibrium state of the system which is,
at the moment when the probe pulse starts, given by the density matrix $\rho(0)$,
which implicitly depends on the pump-probe delay $\tau$.

In the linear-response regime, the probe-induced dipole moment $d_p(t)$ for $t>0$
is expressed as~\cite{PhysRevA.91.033416}
\begin{equation}
\label{Eq:probe_ind}
 d_p(t)=\int\dif t'\mrd\chi(t,t')\mrd e(t'),
\end{equation}
where $\chi(t,t')$ is the nonequilibrium retarded dipole-dipole correlation function
\begin{equation}
\label{Eq:noneq-corr}
 \chi(t,t')=-\frac{\im}{\hbar}\theta(t-t')\mrd\mathrm{Tr}\left(\rho(0)[P(t),P(t')]\right).
\end{equation}
Time dependence in Eq.~\eqref{Eq:noneq-corr} is governed by the Hamiltonian of the system in the absence of
external fields [Eq.~\eqref{Eq:exc_ham_2nd}]
\begin{equation}
\label{Eq:H_field_abs}
 H_{E(t)=0}=H_0+H_\mathrm{e-ph},
\end{equation}
where $H_0$ is the noninteracting Hamiltonian of excitons in the phonon field [the first two terms in Eq.~\eqref{Eq:exc_ham_2nd}],
while $H_\mathrm{e-ph}$ accounts for exciton-phonon interaction [the third term in Eq.~\eqref{Eq:exc_ham_2nd}].
For an ultrashort probe pulse, $e(t)=e_0\delta(t)$, the probe-induced dipole moment assumes the form
\begin{equation}
\label{Eq:probe_ind_ultrashort}
 d_p(t)=e_0\chi(t,0)=e_0\left(-\frac{\im}{\hbar}\right)\mathrm{Tr}\left(\rho(0)[P(t),P(0)]\right).
\end{equation}
Probe pulse tests the possibility of transitions between various exciton states, i.e., it primarily affects carriers.
Therefore, as a reasonable approximation to the full time dependent operator $P(t)$
appearing in Eq.~\eqref{Eq:probe_ind_ultrashort}, operator $P^{(0)}(t)$, evolving according to
the noninteracting Hamiltonian $H_0$ in Eq.~\eqref{Eq:H_field_abs}, may be used. This leads us to the central result for the
probe-induced dipole moment:
\begin{equation}
\label{Eq:probe_ind_central}
 d_p(t)=e_0\left(-\frac{\im}{\hbar}\right)\mathrm{Tr}\left(\rho(0)[P^{(0)}(t),P(0)]\right).
\end{equation}
Deriving the commutator in Eq.~\eqref{Eq:probe_ind_central}, in the expression for $d_p(t)$
we obtain two types of contributions, see Eq.~\eqref{Eq:commutator_t} in Appendix~\ref{Sec:appendix1}.
Contributions of the first type oscillate
at frequencies $\omega_x$ corresponding to
probe-induced transitions between the ground state and exciton state $x$, while those of the second type
oscillate at frequencies $\omega_{\bar x}-\omega_x$ corresponding to probe-induced transitions between
exciton states $\bar x$ and $x$. Here, we focus our attention to the process of
photoinduced absorption (PIA), in which an exciton in state $x$ performs a transition to another state $\bar x$
under the influence of the probe field. Therefore we will further consider only the second type of contributions.

The frequency-dependent transmission coefficient $T(\omega)$ is defined as (we use SI units)
\begin{equation}
\label{Eq:T_def}
 T(\omega)=1+\frac{c\mu_0}{\mathcal{S}\hbar}\mrd\mathrm{Im}\left\{\hbar\omega\mrd\frac{d_p(\omega)}{e(\omega)}\right\},
\end{equation}
where $d_p(\omega)$ and $e(\omega)$ are Fourier transformations of $d_p(t)$ and $e(t)$, respectively, while
$\mathcal{S}$ is the irradiated area of the sample. The differential transmission is given as
\begin{equation}
\label{Eq:DeltaT_def}
 \Delta T(\tau;\omega)=T^\mathrm{neq}(\tau;\omega)-T^\mathrm{eq}(\omega).
\end{equation}
The transmission of a system, which is initially (before the action of the probe) unexcited, is denoted by
$T^\mathrm{eq}(\omega)$. The transmission of a pump-driven system $T^\mathrm{neq}(\tau;\omega)$
depends on the time delay $\tau$ between the pump and the probe through the nonequilibrium density matrix
$\rho(0)$. Since our aim is to study the process of PIA and since $T^\mathrm{eq}(\omega)$ is expected
to reflect only transitions involving the ground state, we will not further consider this term.
After a derivation, the details of which are given in Appendix~\ref{Sec:appendix1},
we obtain the expression for the part of the differential transmission signal $\Delta T_\mathrm{PIA}(\tau;\omega)$
accounting for the PIA:
\begin{widetext}
 \begin{equation}
\label{Eq:signal_pia}
 \begin{split}
  \Delta T_{\mathrm{PIA}}(\tau;\omega)&\propto\mathrm{Im}\left\{
\sum_{xx'}\left((M_x
M^{x}_{x'})^*\frac{\hbar\omega}{\hbar\omega-(\hbar\omega_{x'}
-\hbar\omega_x)+\im\eta}y_{x'}(0)
-M_x
M^{x}_{x'}\frac{\hbar\omega}{\hbar\omega+(\hbar\omega_{x'}
-\hbar\omega_x)+\im\eta}y_{x'}^*(0)\right) \right. + \\ &+ \left.
\sum_{\bar x x x'}\left(
M^x_{x'} M^{\bar x}_x
\frac{\hbar\omega}{\hbar\omega+(\hbar\omega_{x'}-\hbar\omega_x)+\im\eta}
y_{x'}^*(0) y_{\bar x}(0)
-M^{x'}_{x} M^{x}_{\bar x}
\frac{\hbar\omega}{\hbar\omega-(\hbar\omega_{x'}-\hbar\omega_x)+\im\eta} y_{\bar
x}^*(0) y_{x'}(0)\right) \right. + \\ &+ \left.
\sum_{\bar x x x'}\left(
M^x_{x'} M^{\bar x}_x
\frac{\hbar\omega}{\hbar\omega+(\hbar\omega_{x'}-\hbar\omega_x)+\im\eta} \bar
n_{x'\bar x}(0)
-M^{x'}_{x} M^{x}_{\bar x}
\frac{\hbar\omega}{\hbar\omega-(\hbar\omega_{x'}-\hbar\omega_x)+\im\eta} \bar
n_{\bar x x'}(0)\right)
\right\}.
 \end{split}
\end{equation}
\end{widetext}
In the last equation, we have explicitly separated the coherent contributions by
introducing the correlated parts of exciton populations and exciton-exciton coherences
$\bar n_{\bar x x}=n_{\bar x x}-y_{\bar x}^*y_x$
[see also Eq.~\eqref{Eq:incoh_pop_x} defining incoherent exciton populations],
while $\eta$ is a positive parameter effectively accounting for the spectral line broadening.~\cite{EurPhysJB.89.128}
$y_x(0)$ denotes the value of the electronic density matrix $y_x$ at the moment when the probe
pulse starts, and similarly for $\bar n_{\bar x x}(0)$.
The coherences between exciton states and the ground state $y_x(0)$, as well as correlated parts of exciton-exciton
coherences $\bar n_{\bar x x}(0)$ ($\bar x\neq x$),
are expected to approach zero for sufficiently long
time delays between the pump and the probe.~\footnote{Coherences between exciton states and the ground state $y_x$
typically decay on $\lesssim$100-fs time scale after the pump field has vanished, while
exciton-exciton coherences typically decay on ps time scales or longer. Therefore, in our computations,
we expect to see the decay of coherences between exciton states and the ground state, but not of the exciton-exciton coherences,
and time scales on which Eq.~\eqref{Eq:IV_sab} is valid are in principle at least ps or longer.}
In this limit, Eq.~\eqref{Eq:signal_pia} contains only the incoherent exciton populations $\bar n_{xx}$:
\begin{widetext}
\begin{equation}
\label{Eq:IV_sab}
  \Delta T_{\mathrm{PIA}}(\tau;\omega)\propto
\sum_{xx'}|M^x_{x'}|^2\left(-\frac{\eta\cdot\hbar\omega}{(\hbar\omega+(\hbar\omega_{x'}-\hbar\omega_x))^2+\eta^2}
+\frac{\eta\cdot\hbar\omega}{(\hbar\omega-(\hbar\omega_{x'}-\hbar\omega_x))^2+\eta^2}\right)\bar n_{x'x'}(0).
\end{equation}
\end{widetext}
This expression is manifestly negative when it describes probe-induced transitions from exciton state $x'$
to some higher-energy exciton state $x$. The last conclusion is in agreement with the usual experimental
interpretation of pump-probe spectra, where a negative differential transmission signal corresponds either to
PIA or to stimulated emission.~\cite{AdvMater.23.5468}
Our expression [Eq.~\eqref{Eq:signal_pia}] demonstrates, however, that this correspondence
can not be uniquely established in the ultrafast regime, where it is expected that both
coherences between exciton states and the ground state $y_x(0)$ and exciton-exciton coherences
$\bar n_{\bar x x}(0)$ ($\bar x\neq x$), along with incoherent exciton populations $\bar n_{xx}(0)$,
play significant role. This is indeed the case in our numerical computations of pump-probe spectra,
which are presented in the following subsection. For each studied case, we
separately show the total signal [full Eq.~\eqref{Eq:signal_pia}], the $y$-part of the signal
[the first two terms in Eq.~\eqref{Eq:signal_pia}], and the $\bar n$-part of the signal
[the third term in Eq.~\eqref{Eq:signal_pia}]. We note that it would be possible to further separate the $\bar n$-part of the signal
into the contribution stemming from incoherent exciton populations $\bar n_{xx}$ [Eq.~\eqref{Eq:IV_sab}]
and exciton-exciton coherences $\bar n_{\bar x x}$ $(\bar x\neq x)$.
As shown in more detail in Supplemental Material
(Supplemental Fig. 7), the overall $\bar n$-part of the signal is qualitatively very similar
to its contribution stemming from incoherent exciton populations. Therefore, for the simplicity of
further discussion, we may consider the $\bar n$-part of the signal as completely originating from incoherent exciton populations. 

\subsection{Numerical results: ultrafast pump-probe signals}
\label{SSec:num_spectro}
In order to compute pump-probe signals and at the same time keep the numerics manageable,
we extended our model by introducing only one additional
single-electron level both in the donor and in the acceptor and one additional
single-hole level in the donor.
Additional energy levels in the donor and the corresponding bandwidths are extracted from the aforementioned
electronic structure calculation on the infinitely long PCPDTBT polymer.
The additional single-electron level is located at 1160 meV above the single-electron level used in all the calculations
and the bandwidth of the corresponding zone is estimated to be 480 meV.
The additional single-hole level is located at 1130 meV below the single-hole level used in all the calculations
and the bandwidth of the corresponding zone is estimated to be 570 meV.
The additional single-electron level in the acceptor is extracted from an electronic structure calculation
on the C$_{60}$ molecule. The calculation is based on DFT using either LDA or B3LYP exchange-correlation functional
(both choices give similar results) and 6-31G basis set
and was performed using the NWChem package.~\cite{cpc181-1477}
We found that the additional single-electron level lies around 1000 meV above the single-electron level used in all the calculations.
The bandwidth of the corresponding zone is set to 600 meV, see Table~\ref{Tab:model_params}.

In this subsection we assume that the waveform of the pump pulse is
\begin{equation}
\label{Eq:pump_waveform}
 E(t)=E_0\cos(\omega_c t)\exp\left(-\frac{t^2}{\tau_G^2}\right)\theta(t+t_0)\theta(t_0-t),
\end{equation}
where we take $\tau_G=20$ fs and $t_0=50$ fs, while the probe is
\begin{equation}
 \label{Eq:probe_waveform}
 e(t)=e_0\delta(t-(t_0+\tau)),
\end{equation}
with variable pump-probe delay $\tau$.
The intraband dipole matrix elements $d^{\cb\cb}_i,d^{\vb\vb}_i$
in Eq.~\eqref{Eq:P_c_d} are assumed to be equal in the whole system
\begin{equation}
 d^\mathrm{cc}_i=d^\mathrm{vv}_i=d^\mathrm{intra}=\frac{1}{2}d^{\cb\vb}.
\end{equation} 
The positive parameter $\eta$, which effectively accounts for the line broadening, is set to $\eta=50$ meV.
We have checked that variations in $\eta$ do not change the qualitative features of the presented PIA spectra,
see Supplemental Fig. 6.
In actual computations of the signal given in Eq.~\eqref{Eq:signal_pia}, we should remember that the pump pulse finishes
at instant $t_0$, while in Eq.~\eqref{Eq:signal_pia} all the quantities are taken at the moment when the probe starts,
which is now $t_0+\tau$; in other words, $y_x(0)\to y_x(t_0+\tau)$, $\bar n_{\bar x x}(0)\to\bar n_{\bar x x}(t_0+\tau)$
when we compute pump-probe signals using Eq.~\eqref{Eq:signal_pia} and the pump and probe are given by
Eqs.~\eqref{Eq:pump_waveform} and~\eqref{Eq:probe_waveform}, respectively.

\begin{figure}[htbp]
 \begin{center}
  \includegraphics{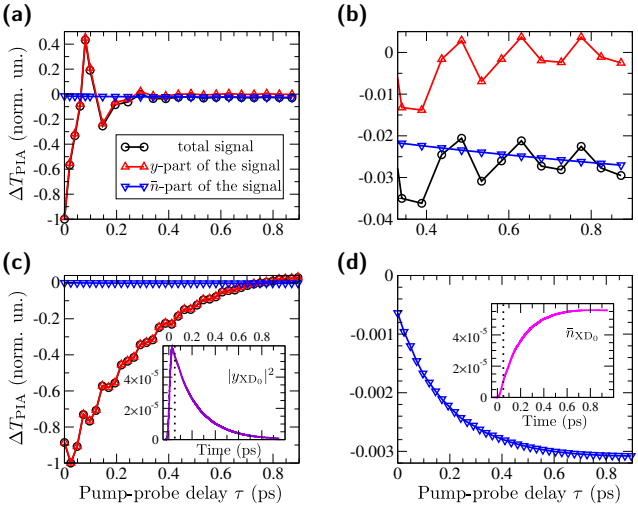}
 \end{center}
 \caption{(Color online) Differential transmission signal $\Delta T_\mathrm{PIA}$ [Eq.~\eqref{Eq:signal_pia}] as a function of the pump-probe delay for:
(a) pump at 1500 meV (826 nm) and probe at 1000 meV (1240 nm) testing PIA dynamics from space-separated states, and
(c) pump resonant with the lowest donor exciton (1210 meV, 1025 nm) and probe at 1130 meV (1096 nm) testing PIA dynamics from donor states.
The inset of (c) shows the coherent exciton population $|y_{\mathrm{XD}_0}|^2$ of the lowest donor state XD$_0$.
(b) The same signal as in (a) at longer pump-probe delays ($>300$ fs).
(d) $\bar n$-part of the signal shown in (c); the inset displays the incoherent exciton population $\bar n_{\mathrm{XD}_0}$ of the lowest
donor state.}
 \label{Fig:fig7}
\end{figure}

In Figs.~\ref{Fig:fig7}(a) and~\ref{Fig:fig7}(b) we show the PIA signal from space-separated states after the excitation by the pump at 1500 meV.
The frequency $\omega$ in Eq.~\eqref{Eq:signal_pia} is set to 1000 meV, which is (for the adopted values of model parameters)
appropriate for observing PIA from space-separated states.
At small pump-probe delays ($\tau\lesssim 300$ fs), we see that the oscillatory features stemming from
coherences between exciton states and the ground state ($y$-part of the signal) dominate the dynamics.
At larger delays, the part originating from established (incoherent) exciton
populations ($\bar n$-part of the signal) prevails, see Fig.~\ref{Fig:fig7}(b), and the shape of the signal resembles the shapes of
signals from space-separated states in Fig. 4(c) of
Ref.~\onlinecite{nmat12-29}. The signal decreases at larger delays,
which correlates very well with the fact that the numbers
of CT and CS excitons increase, see Fig.~\ref{Fig:fig3}(a).
In other words, at larger pump-probe delays,
at which the influence of coherences between exciton states and the ground state is small,
the signal can be unambiguously interpreted in terms of
charge transfer from the donor to the acceptor.

Figures~\ref{Fig:fig7}(c) and~\ref{Fig:fig7}(d) display PIA signal from donor excitons following the pump excitation
at the lowest donor exciton (1210 meV).
The frequency $\omega$ in Eq.~\eqref{Eq:signal_pia} is set to 1130 meV.
The overall signal shape is qualitatively
similar to the shape of donor exciton PIA signal in Fig. 4(a)
of Ref.~\onlinecite{nmat12-29}, but its interpretation is rather different.
While the authors of Ref.~\onlinecite{nmat12-29} suggest that the monotonically
increasing PIA signal from donor excitons reflects their transfer to space-separated
states, our signal predominantly originates from coherences between donor states and the
ground state [$y$-part of the signal in Fig.~\ref{Fig:fig7}(c)].
Furthermore, the shape of the total signal matches very well the decay of the coherent population
of the lowest donor exciton, see the inset of Fig.~\ref{Fig:fig7}(c), while the shape of the $\bar n$-part of the signal corresponds well to the changes in the
incoherent population of the lowest donor state, see the inset of Fig.~\ref{Fig:fig7}(d).
This incoherent population does not decay during our computation: immediately after the pump pulse, it
rises and at longer times it reaches a plateau, which signals that the donor exciton population is "blocked" in the lowest
donor state. The lowest donor exciton is very strongly dipole-coupled to the ground state,
its population comprising around 75\% of the total generated population. Therefore, according to our numerical results,
the observed PIA signal from donor excitons in this case mimics the conversion from coherent
to incoherent exciton population
of the lowest donor state.
This, however, does not necessarily mean that the concomitant charge transfer
is completely absent in this case. Instead, the presence of coherences between exciton states and the ground state,
which dominate the signal for all pump-probe delays we studied, prevents us from attributing the signal
to the population transfer from donor excitons to space-separated states.
The aforementioned conversion from coherent to incoherent exciton population
of the lowest donor state is
rather slow because of the relatively weak coupling between low-lying donor
excitons on the one hand and space-separated
states on the other hand (this weak coupling was also appreciated in
Ref.~\onlinecite{nmat12-29}). On the other hand, pumping well above the lowest donor and space-separated states,
the couplings between these species are stronger and more diverse
than for the pump resonant with the lowest donor exciton; this situation resembles the one encountered
for the excitation condition in Fig. 4(c) of Ref.~\onlinecite{nmat12-29}.

In conclusion, our computations yield spectra which overall agree with experimental spectra,~\cite{nmat12-29}
and we find that the shape of the spectrum in Figs.~\ref{Fig:fig7}(c),(d) originates from the decay of coherences between donor excitons and the ground state,
rather than from transitions from donor excitons to space-separated states.

\section{Discussion and conclusion}
\label{Sec:discuss_conclude}
We studied ultrafast exciton dynamics in a one-dimensional model of a heterointerface.
Even though similar theoretical models have been lately
proposed,~\cite{jpcc.119.7590,jpcc.119.14989} we believe that our theoretical
treatment goes beyond the existing approaches, since it treats both the exciton generation and their further separation on
equal footing and it deals with all the relevant interactions on a fully quantum level. 
Namely, the vast majority of the existing theoretical studies on charge
separation at heterointerfaces does not treat explicitly the interaction with
the electric field which creates
excitons from an initially unexcited system,~\cite{jpcc.119.7590,jpcc.119.14989,PhysRevB.90.115420,PhysRevB.91.041107,PhysRevB.91.201302}
but rather assumes that the exciton has already been generated and then follows its evolution at the interface
between two materials. If we are to explore the possibility of direct optical generation of
space-separated charges, we should certainly monitor the initial process of exciton generation, which we
are able to achieve with the present formalism.
We find that the resonant electronic
coupling between donor and space-separated states
not only enhances transfer from the former to the latter group of states,~\cite{nmat12-29,FD.163.377} but also
opens up a new pathway to obtain space-separated charges: their direct optical generation.~\cite{ADMA:ADMA201402294,jpcl.7.536}
While this mechanism has been proposed on the basis of electronic structure and model Hamiltonian calculations
(which did not include any dynamics),
our study is, to the best of our knowledge, the first to investigate the possibility of direct optical generation of separated charges
studying the ultrafast exciton dynamics at a heterointerface.
We conclude that the largest part of space-separated charges which are present $\sim 100$ fs after the initial photoexcitation are
directly optically generated, contrary to the general belief that they originate from ultrafast transitions from donor excitons.
Although the D/A coupling in our model is restricted to only two nearest sites (labeled by $N-1$ and $N$) in the donor and acceptor,
there are space-separated states which acquire nonzero dipole moment from donor excitons.
The last point was previously highlighted in studies conducted on
two-~\cite{ADMA:ADMA201402294} and three-dimensional~\cite{jpcl.7.536} heterojunction models,
in which the dominant part of the D/A coupling involves more than a single pair of sites.
We thus speculate that the main conclusions of our study would remain valid in a more realistic higher-dimensional model
of a heterointerface.
While there is absorption intensity transfer from donor to space-separated states brought about by their resonant mixing,
the absorption still primarily occurs in the donor part of a heterojunction. Our results show that on ultrafast time scales
the direct optical generation as a source of space-separated carriers is more important than transitions from donor to
space-separated states. This, however, does not mean that initially generated donor excitons do not transform into space-separated states.
They indeed do, see Figs.~\ref{Fig:fig3}(a) and (b), but the characteristic time scale on which populations of space-separated states change 
due to the free-system evolution is longer than 100 fs.

The ultrafast generation of separated charges at heterointerfaces is more pronounced when the electronic coupling between materials is larger or
when the energy overlap region between single-electron states in the donor and acceptor is wider, either by increasing the electronic coupling
in the acceptor or decreasing the LUMO-LUMO offset between the two materials, see Fig.~\ref{Fig:fig2}b.
Our results are therefore in agreement with studies emphasizing the beneficial effects of
larger electronic couplings among materials,~\cite{jpcc.119.7590} charge delocalization,~\cite{PhysRevB.90.115420,jpcc.119.7590,jpcc.119.14989,PhysChemChemPhys.16.20305}
and smaller LUMO-LUMO offset~\cite{ApplPhysLett.88.093511} on charge separation.
We find that strong carrier-phonon interaction suppresses charge separation,
in agreement with previous theoretical studies~\cite{jpcc.119.14989,PhysRevB.91.041107}
in which the effects of variations of carrier-phonon coupling constants have been systematically investigated.
However, changes in the quantities we use to monitor charge separation with variations
of carrier-phonon coupling strength are rather small, which we interpret to be consistent
with the ultrafast direct optical generation of space-separated charges.
Our theoretical treatment of ultrafast exciton dynamics is fully quantum, but it is expected to be valid for not too strong
coupling of excitons to lattice vibrations,
since the phonon branch of the hierarchy is truncated at a finite order, see Sec. I in Supplemental Material.
Results of our mixed quantum/classical approach to exciton dynamics show that
the feedback effect of excitons on the lattice motion, which is expected to be important for stronger exciton-phonon interaction,
is rather small. We therefore expect that more accurate treatment of exciton-phonon interaction is not crucial to describe
heterojunction dynamics on ultrafast time scales. If one wants to treat more accurately strong exciton-phonon interaction
and yet remain in the quantum framework,
other theoretical approaches, such as the one adopted in Ref.~\onlinecite{PhysRevB.91.041107}, have to be employed.

Despite a simplified model of organic semiconductors, our theoretical treatment takes into account all relevant effects.
Consequently, our approach to ultrafast pump-probe experiments
produces results that are in qualitative agreement with experiments and confirms the previously observed dependence of the exciton
dynamics on the excess photon energy.~\cite{nmat12-29}
Our results indicate that the interpretation of ultrafast pump-probe signals is involved, as it is hindered by coherences
(dominantly by those between exciton states and the ground state)
which cannot be neglected on the time scales studied.
Time scales on which coherent features are prominent depend on the excess photon energy.
We find that higher values of the excess photon energy enable faster disappearance of the coherent part of the signal
since they offer diverse transitions between exciton states which make conversion from coherent to incoherent exciton populations faster.
Pumping at the lowest donor exciton, our signal is (at sub-ps pump-probe delays) dominated by its coherent part,
conversion from coherent to incoherent exciton populations is slow, and therefore it cannot be interpreted in terms of exciton population transfer between various states.

\acknowledgments
We gratefully acknowledge the support by the Ministry of Education, Science and Technological
Development of the Republic of Serbia (Project No. ON171017) and European Community FP7
Marie Curie Career Integration Grant (ELECTROMAT), as well as the contribution of the COST Action MP1406.
Numerical computations were performed on the PARADOX supercomputing
facility at the Scientific Computing Laboratory of the Institute of Physics Belgrade.

\begin{widetext}
 \appendix
 \section{Details of the theoretical treatment of pump-probe experiments}
 \label{Sec:appendix1}
The commutator in Eq.~\eqref{Eq:probe_ind_central} is to be evaluated in the nonequilibrium state $\rho(0)$ at the moment when the probe pulse
starts. Therefore, deriving this commutator, only contributions whose expectation values are at most of the second order in the pump field
should be retained. The commutation relations of exciton operators, which are correct up to the second order in the pump field, read as
\begin{equation}
 [X_x,X_{\bar x}^\dagger]=\delta_{x\bar x}-\sum_{\bar x' x'}C^{\bar x' x'}_{\bar x x} X_{\bar x'}^\dagger X_{x'},
\end{equation}
where four-index coefficients $C^{\bar x' x'}_{\bar x x}$ are given as
\begin{equation}
\begin{split}
 C^{\bar x' x'}_{\bar x x}&=\sum_{\substack{\bar j\bar\beta_j\\j\beta_j}}\left(\sum_{i\alpha_i}\psi^{\bar x'*}_{(i\alpha_i)(\bar j\bar\beta_j)}\psi^{x'}_{(i\alpha_i)(j\beta_j)}\right)
\left(\sum_{i\alpha_i}\psi^{\bar x}_{(i\alpha_i)(\bar j\bar\beta_j)}\psi^{x*}_{(i\alpha_i)(j\beta_j)}\right)\\
&+\sum_{\substack{\bar i\bar\alpha_i\\i\alpha_i}}\left(\sum_{j\beta_j}\psi^{\bar x'*}_{(\bar i\bar\alpha_i)(j\beta_j)}\psi^{x'}_{(i\alpha_i)(j\beta_j)}\right)
\left(\sum_{j\beta_j}\psi^{\bar x}_{(\bar i\bar\alpha_i)(j\beta_j)}\psi^{x*}_{(i\alpha_i)(j\beta_j)}\right).
\end{split}
\end{equation}
The final result for the commutator $[P^{(0)}(t),P(0)]$ is
\begin{equation}
\label{Eq:commutator_t}
 \begin{split}
  [P^{(0)}(t),P(0)]&=\sum_{x}|M_x|^2\left(\e^{-\im\omega_x t}-\e^{\im\omega_x t}\right)
-\sum_{\bar{x}_1 x_1}\sum_{xx'}\left(M_x^* M_{x'} C^{\bar{x}_1 x_1}_{x'x}\e^{-\im\omega_x t}-M_x M_{x'}^* C^{\bar{x}_1x_1}_{xx'}\e^{\im\omega_x t}\right)X_{\bar{x}_1}^\dagger X_{x_1}\\
&-\sum_{xx'}\left(M_x M^x_{x'}\right)^*\e^{-\im\omega_x t}X_{x'}+\sum_{xx'} M_x M^x_{x'}\e^{\im\omega_x t}X_{x'}^\dagger\\
&+\sum_{xx'}\left(M_x M^x_{x'}\right)^*\e^{-\im(\omega_{x'}-\omega_x) t}X_{x'}-\sum_{xx'} M_x M^x_{x'}\e^{\im(\omega_{x'}-\omega_x) t}X_{x'}^\dagger\\
&+\sum_{\bar x x x'}M^x_{x'} M^{\bar x}_x \e^{\im(\omega_{x'}-\omega_x)t} X_{x'}^\dagger X_{\bar x}-
\sum_{\bar x x x'}M^{x'}_x M^x_{\bar x} \e^{-\im(\omega_{x'}-\omega_x)t} X_{\bar x}^\dagger X_{x'}.
 \end{split}
\end{equation}
The expectation values (with respect to $\rho(0)$) of the operators appearing in the last equation are simply
the active purely electronic density matrices
of our formalism computed when the probe pulse starts, i.e., $\mathrm{Tr}\left(\rho(0)X_x\right)=y_x(0)$ and
$\mathrm{Tr}\left(\rho(0)X_{\bar x}^\dagger X_x\right)=n_{\bar x x}(0)$.

As already mentioned, in order to study the process of PIA, in Eq.~\eqref{Eq:commutator_t} only terms which oscillate
at differences of two exciton frequencies should be retained. Computing the Fourier transformation of $d_p(t)$
[Eq.~\eqref{Eq:probe_ind_central}], we obtain integrals of the type
\begin{equation}
\label{Eq:fourier}
 \int_0^{+\infty}\dif t\mrd\e^{\im(\omega-\Omega+\im\eta)t}=\frac{\im}{\omega-\Omega+\im\eta},
\end{equation}
where we have introduced a positive infinitesimal parameter $\eta$ to ensure the integral convergence. Physically,
introducing $\eta$ effectively accounts for the line broadening. For simplicity, we assume that only one value of $\eta$
is used in all the integrals of the type~\eqref{Eq:fourier}. Using the computed Fourier transformation $d_p(\omega)$
in Eqs.~\eqref{Eq:T_def} and~\eqref{Eq:DeltaT_def} we obtain the result for $\Delta T_\mathrm{PIA}(\tau;\omega)$
given in Eq.~\eqref{Eq:signal_pia}.
\end{widetext}
\newpage
\bibliography{refs}

\begin{thebibliography}{57}%
\makeatletter
\providecommand \@ifxundefined [1]{%
 \@ifx{#1\undefined}
}%
\providecommand \@ifnum [1]{%
 \ifnum #1\expandafter \@firstoftwo
 \else \expandafter \@secondoftwo
 \fi
}%
\providecommand \@ifx [1]{%
 \ifx #1\expandafter \@firstoftwo
 \else \expandafter \@secondoftwo
 \fi
}%
\providecommand \natexlab [1]{#1}%
\providecommand \enquote  [1]{``#1''}%
\providecommand \bibnamefont  [1]{#1}%
\providecommand \bibfnamefont [1]{#1}%
\providecommand \citenamefont [1]{#1}%
\providecommand \href@noop [0]{\@secondoftwo}%
\providecommand \href [0]{\begingroup \@sanitize@url \@href}%
\providecommand \@href[1]{\@@startlink{#1}\@@href}%
\providecommand \@@href[1]{\endgroup#1\@@endlink}%
\providecommand \@sanitize@url [0]{\catcode `\\12\catcode `\$12\catcode
  `\&12\catcode `\#12\catcode `\^12\catcode `\_12\catcode `\%12\relax}%
\providecommand \@@startlink[1]{}%
\providecommand \@@endlink[0]{}%
\providecommand \url  [0]{\begingroup\@sanitize@url \@url }%
\providecommand \@url [1]{\endgroup\@href {#1}{\urlprefix }}%
\providecommand \urlprefix  [0]{URL }%
\providecommand \Eprint [0]{\href }%
\providecommand \doibase [0]{http://dx.doi.org/}%
\providecommand \selectlanguage [0]{\@gobble}%
\providecommand \bibinfo  [0]{\@secondoftwo}%
\providecommand \bibfield  [0]{\@secondoftwo}%
\providecommand \translation [1]{[#1]}%
\providecommand \BibitemOpen [0]{}%
\providecommand \bibitemStop [0]{}%
\providecommand \bibitemNoStop [0]{.\EOS\space}%
\providecommand \EOS [0]{\spacefactor3000\relax}%
\providecommand \BibitemShut  [1]{\csname bibitem#1\endcsname}%
\let\auto@bib@innerbib\@empty
\bibitem [{\citenamefont {Clarke}\ and\ \citenamefont
  {Durrant}(2010)}]{ChemRev.110.6736}%
  \BibitemOpen
  \bibfield  {author} {\bibinfo {author} {\bibfnamefont {T.~M.}\ \bibnamefont
  {Clarke}}\ and\ \bibinfo {author} {\bibfnamefont {J.~R.}\ \bibnamefont
  {Durrant}},\ }\bibfield  {title} {\enquote {\bibinfo {title} {Charge
  photogeneration in organic solar cells},}\ }\href {\doibase
  10.1021/cr900271s} {\bibfield  {journal} {\bibinfo  {journal} {Chem. Rev.}\
  }\textbf {\bibinfo {volume} {110}},\ \bibinfo {pages} {6736--6767} (\bibinfo
  {year} {2010})}\BibitemShut {NoStop}%
\bibitem [{\citenamefont {Deibel}\ and\ \citenamefont
  {Dyakonov}(2010)}]{RepProgPhys.73.096401}%
  \BibitemOpen
  \bibfield  {author} {\bibinfo {author} {\bibfnamefont {C.}~\bibnamefont
  {Deibel}}\ and\ \bibinfo {author} {\bibfnamefont {V.}~\bibnamefont
  {Dyakonov}},\ }\bibfield  {title} {\enquote {\bibinfo {title}
  {Polymer-fullerene bulk heterojunction solar cells},}\ }\href
  {http://stacks.iop.org/0034-4885/73/i=9/a=096401} {\bibfield  {journal}
  {\bibinfo  {journal} {Rep. Prog. Phys.}\ }\textbf {\bibinfo {volume} {73}},\
  \bibinfo {pages} {096401} (\bibinfo {year} {2010})}\BibitemShut {NoStop}%
\bibitem [{\citenamefont {Gao}\ and\ \citenamefont
  {Ingan{\"a}s}(2014)}]{PCCP.16.20291}%
  \BibitemOpen
  \bibfield  {author} {\bibinfo {author} {\bibfnamefont {F.}~\bibnamefont
  {Gao}}\ and\ \bibinfo {author} {\bibfnamefont {O.}~\bibnamefont
  {Ingan{\"a}s}},\ }\bibfield  {title} {\enquote {\bibinfo {title} {Charge
  generation in polymer-fullerene bulk-heterojunction solar cells},}\ }\href
  {\doibase 10.1039/C4CP01814A} {\bibfield  {journal} {\bibinfo  {journal}
  {Phys. Chem. Chem. Phys.}\ }\textbf {\bibinfo {volume} {16}},\ \bibinfo
  {pages} {20291--20304} (\bibinfo {year} {2014})}\BibitemShut {NoStop}%
\bibitem [{\citenamefont {Zhugayevych}\ and\ \citenamefont
  {Tretiak}(2015)}]{AnnRevPhysChem.66.305}%
  \BibitemOpen
  \bibfield  {author} {\bibinfo {author} {\bibfnamefont {A.}~\bibnamefont
  {Zhugayevych}}\ and\ \bibinfo {author} {\bibfnamefont {S.}~\bibnamefont
  {Tretiak}},\ }\bibfield  {title} {\enquote {\bibinfo {title} {Theoretical
  description of structural and electronic properties of organic photovoltaic
  materials},}\ }\href {\doibase 10.1146/annurev-physchem-040214-121440}
  {\bibfield  {journal} {\bibinfo  {journal} {Annu. Rev. Phys. Chem.}\ }\textbf
  {\bibinfo {volume} {66}},\ \bibinfo {pages} {305--330} (\bibinfo {year}
  {2015})}\BibitemShut {NoStop}%
\bibitem [{\citenamefont {B{\"a}ssler}\ and\ \citenamefont
  {K{\"o}hler}(2015)}]{PCCP.17.28451}%
  \BibitemOpen
  \bibfield  {author} {\bibinfo {author} {\bibfnamefont {H.}~\bibnamefont
  {B{\"a}ssler}}\ and\ \bibinfo {author} {\bibfnamefont {A.}~\bibnamefont
  {K{\"o}hler}},\ }\bibfield  {title} {\enquote {\bibinfo {title} {"{H}ot or
  cold": how do charge transfer states at the donor-acceptor interface of an
  organic solar cell dissociate?}}\ }\href {\doibase 10.1039/C5CP04110D}
  {\bibfield  {journal} {\bibinfo  {journal} {Phys. Chem. Chem. Phys.}\
  }\textbf {\bibinfo {volume} {17}},\ \bibinfo {pages} {28451--28462} (\bibinfo
  {year} {2015})}\BibitemShut {NoStop}%
\bibitem [{\citenamefont {Br{\'{e}}das}\ \emph {et~al.}(2009)\citenamefont
  {Br{\'{e}}das}, \citenamefont {Norton}, \citenamefont {Cornil},\ and\
  \citenamefont {Coropceanu}}]{AcChemRes.42.1691}%
  \BibitemOpen
  \bibfield  {author} {\bibinfo {author} {\bibfnamefont {J.~L.}\ \bibnamefont
  {Br{\'{e}}das}}, \bibinfo {author} {\bibfnamefont {J.~E.}\ \bibnamefont
  {Norton}}, \bibinfo {author} {\bibfnamefont {J.}~\bibnamefont {Cornil}}, \
  and\ \bibinfo {author} {\bibfnamefont {V.}~\bibnamefont {Coropceanu}},\
  }\bibfield  {title} {\enquote {\bibinfo {title} {Molecular understanding of
  organic solar cells: the challenges},}\ }\href {\doibase 10.1021/ar900099h}
  {\bibfield  {journal} {\bibinfo  {journal} {Acc. Chem. Res.}\ }\textbf
  {\bibinfo {volume} {42}},\ \bibinfo {pages} {1691--1699} (\bibinfo {year}
  {2009})}\BibitemShut {NoStop}%
\bibitem [{\citenamefont {Grancini}\ \emph {et~al.}(2013)\citenamefont
  {Grancini}, \citenamefont {Maiuri}, \citenamefont {Fazzi}, \citenamefont
  {Petrozza}, \citenamefont {Egelhaaf}, \citenamefont {Brida}, \citenamefont
  {Cerullo},\ and\ \citenamefont {Lanzani}}]{nmat12-29}%
  \BibitemOpen
  \bibfield  {author} {\bibinfo {author} {\bibfnamefont {G.}~\bibnamefont
  {Grancini}}, \bibinfo {author} {\bibfnamefont {M.}~\bibnamefont {Maiuri}},
  \bibinfo {author} {\bibfnamefont {D.}~\bibnamefont {Fazzi}}, \bibinfo
  {author} {\bibfnamefont {A.}~\bibnamefont {Petrozza}}, \bibinfo {author}
  {\bibfnamefont {H-J.}\ \bibnamefont {Egelhaaf}}, \bibinfo {author}
  {\bibfnamefont {D.}~\bibnamefont {Brida}}, \bibinfo {author} {\bibfnamefont
  {G.}~\bibnamefont {Cerullo}}, \ and\ \bibinfo {author} {\bibfnamefont
  {G.}~\bibnamefont {Lanzani}},\ }\bibfield  {title} {\enquote {\bibinfo
  {title} {Hot exciton dissociation in polymer solar cells},}\ }\href
  {http://www.nature.com/nmat/journal/v12/n1/full/nmat3502.html} {\bibfield
  {journal} {\bibinfo  {journal} {Nat. Mater.}\ }\textbf {\bibinfo {volume}
  {12}},\ \bibinfo {pages} {29--33} (\bibinfo {year} {2013})}\BibitemShut
  {NoStop}%
\bibitem [{\citenamefont {Jailaubekov}\ \emph {et~al.}(2013)\citenamefont
  {Jailaubekov}, \citenamefont {Willard}, \citenamefont {Tritsch},
  \citenamefont {Chan}, \citenamefont {Sai}, \citenamefont {Gearba},
  \citenamefont {Kaake}, \citenamefont {Williams}, \citenamefont {Leung},
  \citenamefont {Rossky},\ and\ \citenamefont {Zhu}}]{nmat12-66}%
  \BibitemOpen
  \bibfield  {author} {\bibinfo {author} {\bibfnamefont {A.~E.}\ \bibnamefont
  {Jailaubekov}}, \bibinfo {author} {\bibfnamefont {A.~P.}\ \bibnamefont
  {Willard}}, \bibinfo {author} {\bibfnamefont {J.~R.}\ \bibnamefont
  {Tritsch}}, \bibinfo {author} {\bibfnamefont {W.-L.}\ \bibnamefont {Chan}},
  \bibinfo {author} {\bibfnamefont {N.}~\bibnamefont {Sai}}, \bibinfo {author}
  {\bibfnamefont {R.}~\bibnamefont {Gearba}}, \bibinfo {author} {\bibfnamefont
  {L.~G.}\ \bibnamefont {Kaake}}, \bibinfo {author} {\bibfnamefont {K.~J.}\
  \bibnamefont {Williams}}, \bibinfo {author} {\bibfnamefont {K.}~\bibnamefont
  {Leung}}, \bibinfo {author} {\bibfnamefont {P.~J.}\ \bibnamefont {Rossky}}, \
  and\ \bibinfo {author} {\bibfnamefont {X-Y.}\ \bibnamefont {Zhu}},\
  }\bibfield  {title} {\enquote {\bibinfo {title} {Hot charge-transfer excitons
  set the time limit for charge separation at donor/acceptor interfaces in
  organic photovoltaics},}\ }\href
  {http://www.nature.com/nmat/journal/v12/n1/full/nmat3500.html} {\bibfield
  {journal} {\bibinfo  {journal} {Nat. Mater.}\ }\textbf {\bibinfo {volume}
  {12}},\ \bibinfo {pages} {66--73} (\bibinfo {year} {2013})}\BibitemShut
  {NoStop}%
\bibitem [{\citenamefont {G{\' e}linas}\ \emph {et~al.}(2014)\citenamefont
  {G{\' e}linas}, \citenamefont {Rao}, \citenamefont {Kumar}, \citenamefont
  {Smith}, \citenamefont {Chin}, \citenamefont {Clark}, \citenamefont {van~der
  Poll}, \citenamefont {Bazan},\ and\ \citenamefont {Friend}}]{science343-512}%
  \BibitemOpen
  \bibfield  {author} {\bibinfo {author} {\bibfnamefont {S.}~\bibnamefont {G{\'
  e}linas}}, \bibinfo {author} {\bibfnamefont {A.}~\bibnamefont {Rao}},
  \bibinfo {author} {\bibfnamefont {A.}~\bibnamefont {Kumar}}, \bibinfo
  {author} {\bibfnamefont {S.~L.}\ \bibnamefont {Smith}}, \bibinfo {author}
  {\bibfnamefont {A.~W.}\ \bibnamefont {Chin}}, \bibinfo {author}
  {\bibfnamefont {J.}~\bibnamefont {Clark}}, \bibinfo {author} {\bibfnamefont
  {T.~S.}\ \bibnamefont {van~der Poll}}, \bibinfo {author} {\bibfnamefont
  {G.~C.}\ \bibnamefont {Bazan}}, \ and\ \bibinfo {author} {\bibfnamefont
  {R.~H.}\ \bibnamefont {Friend}},\ }\bibfield  {title} {\enquote {\bibinfo
  {title} {Ultrafast long-range charge separation in organic semiconductor
  photovoltaic diodes},}\ }\href {\doibase 10.1126/science.1246249} {\bibfield
  {journal} {\bibinfo  {journal} {Science}\ }\textbf {\bibinfo {volume}
  {343}},\ \bibinfo {pages} {512--516} (\bibinfo {year} {2014})}\BibitemShut
  {NoStop}%
\bibitem [{\citenamefont {Paraecattil}\ and\ \citenamefont
  {Banerji}(2014)}]{JAmChemSoc.136.1472}%
  \BibitemOpen
  \bibfield  {author} {\bibinfo {author} {\bibfnamefont {A.~A.}\ \bibnamefont
  {Paraecattil}}\ and\ \bibinfo {author} {\bibfnamefont {N.}~\bibnamefont
  {Banerji}},\ }\bibfield  {title} {\enquote {\bibinfo {title} {Charge
  separation pathways in a highly efficient polymer:fullerene solar cell
  material},}\ }\href {http://dx.doi.org/10.1021/ja410340g} {\bibfield
  {journal} {\bibinfo  {journal} {J. Am. Chem. Soc.}\ }\textbf {\bibinfo
  {volume} {136}},\ \bibinfo {pages} {1472--1482} (\bibinfo {year}
  {2014})}\BibitemShut {NoStop}%
\bibitem [{\citenamefont {Cowan}\ \emph {et~al.}(2012)\citenamefont {Cowan},
  \citenamefont {Banerji}, \citenamefont {Leong},\ and\ \citenamefont
  {Heeger}}]{afm22-1116}%
  \BibitemOpen
  \bibfield  {author} {\bibinfo {author} {\bibfnamefont {S.~R.}\ \bibnamefont
  {Cowan}}, \bibinfo {author} {\bibfnamefont {N.}~\bibnamefont {Banerji}},
  \bibinfo {author} {\bibfnamefont {W.~L.}\ \bibnamefont {Leong}}, \ and\
  \bibinfo {author} {\bibfnamefont {A.~J.}\ \bibnamefont {Heeger}},\ }\bibfield
   {title} {\enquote {\bibinfo {title} {Charge formation, recombination, and
  sweep-out dynamics in organic solar cells},}\ }\href
  {http://dx.doi.org/10.1002/adfm.201101632} {\bibfield  {journal} {\bibinfo
  {journal} {Adv. Funct. Mater.}\ }\textbf {\bibinfo {volume} {22}},\ \bibinfo
  {pages} {1116--1128} (\bibinfo {year} {2012})}\BibitemShut {NoStop}%
\bibitem [{\citenamefont {Deibel}\ \emph {et~al.}(2010)\citenamefont {Deibel},
  \citenamefont {Strobel},\ and\ \citenamefont
  {Dyakonov}}]{ADMA:ADMA201000376}%
  \BibitemOpen
  \bibfield  {author} {\bibinfo {author} {\bibfnamefont {C.}~\bibnamefont
  {Deibel}}, \bibinfo {author} {\bibfnamefont {T.}~\bibnamefont {Strobel}}, \
  and\ \bibinfo {author} {\bibfnamefont {V.}~\bibnamefont {Dyakonov}},\
  }\bibfield  {title} {\enquote {\bibinfo {title} {Role of the charge transfer
  state in organic donor-acceptor solar cells},}\ }\href {\doibase
  10.1002/adma.201000376} {\bibfield  {journal} {\bibinfo  {journal} {Adv.
  Mater.}\ }\textbf {\bibinfo {volume} {22}},\ \bibinfo {pages} {4097--4111}
  (\bibinfo {year} {2010})}\BibitemShut {NoStop}%
\bibitem [{\citenamefont {Bakulin}\ \emph {et~al.}(2012)\citenamefont
  {Bakulin}, \citenamefont {Rao}, \citenamefont {Pavelyev}, \citenamefont {van
  Loosdrecht}, \citenamefont {Pshenichnikov}, \citenamefont {Niedzialek},
  \citenamefont {Cornil}, \citenamefont {Beljonne},\ and\ \citenamefont
  {Friend}}]{science335-1340}%
  \BibitemOpen
  \bibfield  {author} {\bibinfo {author} {\bibfnamefont {A.~A.}\ \bibnamefont
  {Bakulin}}, \bibinfo {author} {\bibfnamefont {A.}~\bibnamefont {Rao}},
  \bibinfo {author} {\bibfnamefont {V.~G.}\ \bibnamefont {Pavelyev}}, \bibinfo
  {author} {\bibfnamefont {P.~H.~M.}\ \bibnamefont {van Loosdrecht}}, \bibinfo
  {author} {\bibfnamefont {M.~S.}\ \bibnamefont {Pshenichnikov}}, \bibinfo
  {author} {\bibfnamefont {D.}~\bibnamefont {Niedzialek}}, \bibinfo {author}
  {\bibfnamefont {J.}~\bibnamefont {Cornil}}, \bibinfo {author} {\bibfnamefont
  {D.}~\bibnamefont {Beljonne}}, \ and\ \bibinfo {author} {\bibfnamefont
  {R.~H.}\ \bibnamefont {Friend}},\ }\bibfield  {title} {\enquote {\bibinfo
  {title} {The role of driving energy and delocalized states for charge
  separation in organic semiconductors},}\ }\href {\doibase
  10.1126/science.1217745} {\bibfield  {journal} {\bibinfo  {journal}
  {Science}\ }\textbf {\bibinfo {volume} {335}},\ \bibinfo {pages} {1340--1344}
  (\bibinfo {year} {2012})}\BibitemShut {NoStop}%
\bibitem [{\citenamefont {Chen}\ \emph {et~al.}(2013)\citenamefont {Chen},
  \citenamefont {Barker}, \citenamefont {Reish}, \citenamefont {Gordon},\ and\
  \citenamefont {Hodgkiss}}]{jacs.135.18502}%
  \BibitemOpen
  \bibfield  {author} {\bibinfo {author} {\bibfnamefont {K.}~\bibnamefont
  {Chen}}, \bibinfo {author} {\bibfnamefont {A.~J.}\ \bibnamefont {Barker}},
  \bibinfo {author} {\bibfnamefont {M.~E.}\ \bibnamefont {Reish}}, \bibinfo
  {author} {\bibfnamefont {K.~C.}\ \bibnamefont {Gordon}}, \ and\ \bibinfo
  {author} {\bibfnamefont {J.~M.}\ \bibnamefont {Hodgkiss}},\ }\bibfield
  {title} {\enquote {\bibinfo {title} {Broadband ultrafast photoluminescence
  spectroscopy resolves charge photogeneration via delocalized hot excitons in
  polymer:fullerene photovoltaic blends},}\ }\href {\doibase 10.1021/ja408235h}
  {\bibfield  {journal} {\bibinfo  {journal} {J. Am. Chem. Soc.}\ }\textbf
  {\bibinfo {volume} {135}},\ \bibinfo {pages} {18502--18512} (\bibinfo {year}
  {2013})}\BibitemShut {NoStop}%
\bibitem [{\citenamefont {Troisi}(2013)}]{FD.163.377}%
  \BibitemOpen
  \bibfield  {author} {\bibinfo {author} {\bibfnamefont {A.}~\bibnamefont
  {Troisi}},\ }\bibfield  {title} {\enquote {\bibinfo {title} {How quasi-free
  holes and electrons are generated in organic photovoltaic interfaces},}\
  }\href {\doibase 10.1039/C3FD20142B} {\bibfield  {journal} {\bibinfo
  {journal} {Faraday Discuss.}\ }\textbf {\bibinfo {volume} {163}},\ \bibinfo
  {pages} {377--392} (\bibinfo {year} {2013})}\BibitemShut {NoStop}%
\bibitem [{\citenamefont {V\'azquez}\ and\ \citenamefont
  {Troisi}(2013)}]{PhysRevB.88.205304}%
  \BibitemOpen
  \bibfield  {author} {\bibinfo {author} {\bibfnamefont {H.}~\bibnamefont
  {V\'azquez}}\ and\ \bibinfo {author} {\bibfnamefont {A.}~\bibnamefont
  {Troisi}},\ }\bibfield  {title} {\enquote {\bibinfo {title} {Calculation of
  rates of exciton dissociation into hot charge-transfer states in model
  organic photovoltaic interfaces},}\ }\href {\doibase
  10.1103/PhysRevB.88.205304} {\bibfield  {journal} {\bibinfo  {journal} {Phys.
  Rev. B}\ }\textbf {\bibinfo {volume} {88}},\ \bibinfo {pages} {205304}
  (\bibinfo {year} {2013})}\BibitemShut {NoStop}%
\bibitem [{\citenamefont {Sun}\ and\ \citenamefont
  {Stafstr{\"o}m}(2014)}]{PhysRevB.90.115420}%
  \BibitemOpen
  \bibfield  {author} {\bibinfo {author} {\bibfnamefont {Z.}~\bibnamefont
  {Sun}}\ and\ \bibinfo {author} {\bibfnamefont {S.}~\bibnamefont
  {Stafstr{\"o}m}},\ }\bibfield  {title} {\enquote {\bibinfo {title} {Dynamics
  of charge separation at an organic donor-acceptor interface},}\ }\href
  {\doibase 10.1103/PhysRevB.90.115420} {\bibfield  {journal} {\bibinfo
  {journal} {Phys. Rev. B}\ }\textbf {\bibinfo {volume} {90}},\ \bibinfo
  {pages} {115420} (\bibinfo {year} {2014})}\BibitemShut {NoStop}%
\bibitem [{\citenamefont {Nan}\ \emph {et~al.}(2015)\citenamefont {Nan},
  \citenamefont {Zhang},\ and\ \citenamefont {Lu}}]{JPhysChemC.119.15028}%
  \BibitemOpen
  \bibfield  {author} {\bibinfo {author} {\bibfnamefont {G.}~\bibnamefont
  {Nan}}, \bibinfo {author} {\bibfnamefont {X.}~\bibnamefont {Zhang}}, \ and\
  \bibinfo {author} {\bibfnamefont {G.}~\bibnamefont {Lu}},\ }\bibfield
  {title} {\enquote {\bibinfo {title} {Do "hot" charge-transfer excitons
  promote free carrier generation in organic photovoltaics?}}\ }\href@noop {}
  {\bibfield  {journal} {\bibinfo  {journal} {J. Phys. Chem. C}\ }\textbf
  {\bibinfo {volume} {119}},\ \bibinfo {pages} {15028--15035} (\bibinfo {year}
  {2015})}\BibitemShut {NoStop}%
\bibitem [{\citenamefont {Smith}\ and\ \citenamefont
  {Chin}(2015)}]{PhysRevB.91.201302}%
  \BibitemOpen
  \bibfield  {author} {\bibinfo {author} {\bibfnamefont {S.~L.}\ \bibnamefont
  {Smith}}\ and\ \bibinfo {author} {\bibfnamefont {A.~W.}\ \bibnamefont
  {Chin}},\ }\bibfield  {title} {\enquote {\bibinfo {title} {Phonon-assisted
  ultrafast charge separation in the {P}{C}{B}{M} band structure},}\ }\href
  {\doibase 10.1103/PhysRevB.91.201302} {\bibfield  {journal} {\bibinfo
  {journal} {Phys. Rev. B}\ }\textbf {\bibinfo {volume} {91}},\ \bibinfo
  {pages} {201302} (\bibinfo {year} {2015})}\BibitemShut {NoStop}%
\bibitem [{\citenamefont {Tamura}\ and\ \citenamefont
  {Burghardt}(2013)}]{jacs.135.16364}%
  \BibitemOpen
  \bibfield  {author} {\bibinfo {author} {\bibfnamefont {H.}~\bibnamefont
  {Tamura}}\ and\ \bibinfo {author} {\bibfnamefont {I.}~\bibnamefont
  {Burghardt}},\ }\bibfield  {title} {\enquote {\bibinfo {title} {Ultrafast
  charge separation in organic photovoltaics enhanced by charge delocalization
  and vibronically hot exciton dissociation},}\ }\href {\doibase
  10.1021/ja4093874} {\bibfield  {journal} {\bibinfo  {journal} {J. Am. Chem.
  Soc.}\ }\textbf {\bibinfo {volume} {135}},\ \bibinfo {pages} {16364--16367}
  (\bibinfo {year} {2013})}\BibitemShut {NoStop}%
\bibitem [{\citenamefont {Smith}\ and\ \citenamefont
  {Chin}(2014)}]{PhysChemChemPhys.16.20305}%
  \BibitemOpen
  \bibfield  {author} {\bibinfo {author} {\bibfnamefont {S.~L.}\ \bibnamefont
  {Smith}}\ and\ \bibinfo {author} {\bibfnamefont {A.~W.}\ \bibnamefont
  {Chin}},\ }\bibfield  {title} {\enquote {\bibinfo {title} {Ultrafast charge
  separation and nongeminate electron-hole recombination in organic
  photovoltaics},}\ }\href {http://dx.doi.org/10.1039/C4CP01791A} {\bibfield
  {journal} {\bibinfo  {journal} {Phys. Chem. Chem. Phys.}\ }\textbf {\bibinfo
  {volume} {16}},\ \bibinfo {pages} {20305--20309} (\bibinfo {year}
  {2014})}\BibitemShut {NoStop}%
\bibitem [{\citenamefont {Vandewal}\ \emph {et~al.}(2014)\citenamefont
  {Vandewal}, \citenamefont {Albrecht}, \citenamefont {Hoke}, \citenamefont
  {Graham}, \citenamefont {Widmer}, \citenamefont {Douglas}, \citenamefont
  {Schubert}, \citenamefont {Mateker}, \citenamefont {Bloking}, \citenamefont
  {Burkhard}, \citenamefont {Sellinger}, \citenamefont {Fr{\'e}chet},
  \citenamefont {Amassian}, \citenamefont {Riede}, \citenamefont {McGehee},
  \citenamefont {Neher},\ and\ \citenamefont {Salleo}}]{nmat13-63}%
  \BibitemOpen
  \bibfield  {author} {\bibinfo {author} {\bibfnamefont {K.}~\bibnamefont
  {Vandewal}}, \bibinfo {author} {\bibfnamefont {S.}~\bibnamefont {Albrecht}},
  \bibinfo {author} {\bibfnamefont {E.~T.}\ \bibnamefont {Hoke}}, \bibinfo
  {author} {\bibfnamefont {K.~R.}\ \bibnamefont {Graham}}, \bibinfo {author}
  {\bibfnamefont {J.}~\bibnamefont {Widmer}}, \bibinfo {author} {\bibfnamefont
  {J.~D.}\ \bibnamefont {Douglas}}, \bibinfo {author} {\bibfnamefont
  {M.}~\bibnamefont {Schubert}}, \bibinfo {author} {\bibfnamefont {W.~R.}\
  \bibnamefont {Mateker}}, \bibinfo {author} {\bibfnamefont {J.~T.}\
  \bibnamefont {Bloking}}, \bibinfo {author} {\bibfnamefont {G.~F.}\
  \bibnamefont {Burkhard}}, \bibinfo {author} {\bibfnamefont {A.}~\bibnamefont
  {Sellinger}}, \bibinfo {author} {\bibfnamefont {J.~M.~J.}\ \bibnamefont
  {Fr{\'e}chet}}, \bibinfo {author} {\bibfnamefont {A.}~\bibnamefont
  {Amassian}}, \bibinfo {author} {\bibfnamefont {M.~K.}\ \bibnamefont {Riede}},
  \bibinfo {author} {\bibfnamefont {M.~D.}\ \bibnamefont {McGehee}}, \bibinfo
  {author} {\bibfnamefont {D.}~\bibnamefont {Neher}}, \ and\ \bibinfo {author}
  {\bibfnamefont {A.}~\bibnamefont {Salleo}},\ }\bibfield  {title} {\enquote
  {\bibinfo {title} {Efficient charge generation by relaxed charge-transfer
  states at organic interfaces},}\ }\href {http://dx.doi.org/10.1038/nmat3807}
  {\bibfield  {journal} {\bibinfo  {journal} {Nat. Mater.}\ }\textbf {\bibinfo
  {volume} {13}},\ \bibinfo {pages} {63--68} (\bibinfo {year}
  {2014})}\BibitemShut {NoStop}%
\bibitem [{\citenamefont {Bittner}\ and\ \citenamefont
  {Silva}(2014)}]{ncomms5-3119}%
  \BibitemOpen
  \bibfield  {author} {\bibinfo {author} {\bibfnamefont {E.~R.}\ \bibnamefont
  {Bittner}}\ and\ \bibinfo {author} {\bibfnamefont {C.}~\bibnamefont
  {Silva}},\ }\bibfield  {title} {\enquote {\bibinfo {title} {Noise-induced
  quantum coherence drives photo-carrier generation dynamics at polymeric
  semiconductor heterojunctions},}\ }\href
  {http://dx.doi.org/10.1038/ncomms4119} {\bibfield  {journal} {\bibinfo
  {journal} {Nat. Commun.}\ }\textbf {\bibinfo {volume} {5}},\ \bibinfo {pages}
  {3119} (\bibinfo {year} {2014})}\BibitemShut {NoStop}%
\bibitem [{\citenamefont {Savoie}\ \emph {et~al.}(2014)\citenamefont {Savoie},
  \citenamefont {Rao}, \citenamefont {Bakulin}, \citenamefont {Gelinas},
  \citenamefont {Movaghar}, \citenamefont {Friend}, \citenamefont {Marks},\
  and\ \citenamefont {Ratner}}]{jacs.136.2876}%
  \BibitemOpen
  \bibfield  {author} {\bibinfo {author} {\bibfnamefont {B.~M.}\ \bibnamefont
  {Savoie}}, \bibinfo {author} {\bibfnamefont {A.}~\bibnamefont {Rao}},
  \bibinfo {author} {\bibfnamefont {A.~A.}\ \bibnamefont {Bakulin}}, \bibinfo
  {author} {\bibfnamefont {S.}~\bibnamefont {Gelinas}}, \bibinfo {author}
  {\bibfnamefont {B.}~\bibnamefont {Movaghar}}, \bibinfo {author}
  {\bibfnamefont {R.~H.}\ \bibnamefont {Friend}}, \bibinfo {author}
  {\bibfnamefont {T.~J.}\ \bibnamefont {Marks}}, \ and\ \bibinfo {author}
  {\bibfnamefont {M.~A.}\ \bibnamefont {Ratner}},\ }\bibfield  {title}
  {\enquote {\bibinfo {title} {Unequal partnership: Asymmetric roles of
  polymeric donor and fullerene acceptor in generating free charge},}\ }\href
  {\doibase 10.1021/ja411859m} {\bibfield  {journal} {\bibinfo  {journal} {J.
  Am. Chem. Soc.}\ }\textbf {\bibinfo {volume} {136}},\ \bibinfo {pages}
  {2876--2884} (\bibinfo {year} {2014})}\BibitemShut {NoStop}%
\bibitem [{\citenamefont {Ma}\ and\ \citenamefont
  {Troisi}(2014)}]{ADMA:ADMA201402294}%
  \BibitemOpen
  \bibfield  {author} {\bibinfo {author} {\bibfnamefont {H.}~\bibnamefont
  {Ma}}\ and\ \bibinfo {author} {\bibfnamefont {A.}~\bibnamefont {Troisi}},\
  }\bibfield  {title} {\enquote {\bibinfo {title} {Direct optical generation of
  long-range charge-transfer states in organic photovoltaics},}\ }\href
  {\doibase 10.1002/adma.201402294} {\bibfield  {journal} {\bibinfo  {journal}
  {Adv. Mater.}\ }\textbf {\bibinfo {volume} {26}},\ \bibinfo {pages}
  {6163--6167} (\bibinfo {year} {2014})}\BibitemShut {NoStop}%
\bibitem [{\citenamefont {D'Avino}\ \emph {et~al.}(2016)\citenamefont
  {D'Avino}, \citenamefont {Muccioli}, \citenamefont {Olivier},\ and\
  \citenamefont {Beljonne}}]{jpcl.7.536}%
  \BibitemOpen
  \bibfield  {author} {\bibinfo {author} {\bibfnamefont {G.}~\bibnamefont
  {D'Avino}}, \bibinfo {author} {\bibfnamefont {L.}~\bibnamefont {Muccioli}},
  \bibinfo {author} {\bibfnamefont {Y.}~\bibnamefont {Olivier}}, \ and\
  \bibinfo {author} {\bibfnamefont {D.}~\bibnamefont {Beljonne}},\ }\bibfield
  {title} {\enquote {\bibinfo {title} {Charge separation and recombination at
  polymer-fullerene heterojunctions: delocalization and hybridization
  effects},}\ }\href {\doibase 10.1021/acs.jpclett.5b02680} {\bibfield
  {journal} {\bibinfo  {journal} {J. Phys. Chem. Lett.}\ }\textbf {\bibinfo
  {volume} {7}},\ \bibinfo {pages} {536--540} (\bibinfo {year}
  {2016})}\BibitemShut {NoStop}%
\bibitem [{\citenamefont {Axt}\ and\ \citenamefont
  {Stahl}(1994)}]{ZPhysB.93.195}%
  \BibitemOpen
  \bibfield  {author} {\bibinfo {author} {\bibfnamefont {V.M.}\ \bibnamefont
  {Axt}}\ and\ \bibinfo {author} {\bibfnamefont {A.}~\bibnamefont {Stahl}},\
  }\bibfield  {title} {\enquote {\bibinfo {title} {A dynamics-controlled
  truncation scheme for the hierarchy of density matrices in semiconductor
  optics},}\ }\href {\doibase 10.1007/BF01316963} {\bibfield  {journal}
  {\bibinfo  {journal} {Z. Phys. B}\ }\textbf {\bibinfo {volume} {93}},\
  \bibinfo {pages} {195--204} (\bibinfo {year} {1994})}\BibitemShut {NoStop}%
\bibitem [{\citenamefont {Axt}\ \emph {et~al.}(1996)\citenamefont {Axt},
  \citenamefont {Victor},\ and\ \citenamefont {Stahl}}]{PhysRevB.53.7244}%
  \BibitemOpen
  \bibfield  {author} {\bibinfo {author} {\bibfnamefont {V.~M.}\ \bibnamefont
  {Axt}}, \bibinfo {author} {\bibfnamefont {K.}~\bibnamefont {Victor}}, \ and\
  \bibinfo {author} {\bibfnamefont {A.}~\bibnamefont {Stahl}},\ }\bibfield
  {title} {\enquote {\bibinfo {title} {Influence of a phonon bath on the
  hierarchy of electronic densities in an optically excited semiconductor},}\
  }\href {\doibase 10.1103/PhysRevB.53.7244} {\bibfield  {journal} {\bibinfo
  {journal} {Phys. Rev. B}\ }\textbf {\bibinfo {volume} {53}},\ \bibinfo
  {pages} {7244--7258} (\bibinfo {year} {1996})}\BibitemShut {NoStop}%
\bibitem [{\citenamefont {Axt}\ and\ \citenamefont
  {Mukamel}(1998)}]{RevModPhys.70.145}%
  \BibitemOpen
  \bibfield  {author} {\bibinfo {author} {\bibfnamefont {V.~M.}\ \bibnamefont
  {Axt}}\ and\ \bibinfo {author} {\bibfnamefont {S.}~\bibnamefont {Mukamel}},\
  }\bibfield  {title} {\enquote {\bibinfo {title} {Nonlinear optics of
  semiconductor and molecular nanostructures; a common perspective},}\ }\href
  {\doibase 10.1103/RevModPhys.70.145} {\bibfield  {journal} {\bibinfo
  {journal} {Rev. Mod. Phys.}\ }\textbf {\bibinfo {volume} {70}},\ \bibinfo
  {pages} {145--174} (\bibinfo {year} {1998})}\BibitemShut {NoStop}%
\bibitem [{\citenamefont {Jankovi\ifmmode~\acute{c}\else \'{c}\fi{}}\ and\
  \citenamefont {Vukmirovi\ifmmode~\acute{c}\else
  \'{c}\fi{}}(2015)}]{PhysRevB.92.235208}%
  \BibitemOpen
  \bibfield  {author} {\bibinfo {author} {\bibfnamefont {V.}~\bibnamefont
  {Jankovi\ifmmode~\acute{c}\else \'{c}\fi{}}}\ and\ \bibinfo {author}
  {\bibfnamefont {N.}~\bibnamefont {Vukmirovi\ifmmode~\acute{c}\else
  \'{c}\fi{}}},\ }\bibfield  {title} {\enquote {\bibinfo {title} {Dynamics of
  exciton formation and relaxation in photoexcited semiconductors},}\ }\href
  {\doibase 10.1103/PhysRevB.92.235208} {\bibfield  {journal} {\bibinfo
  {journal} {Phys. Rev. B}\ }\textbf {\bibinfo {volume} {92}},\ \bibinfo
  {pages} {235208} (\bibinfo {year} {2015})}\BibitemShut {NoStop}%
\bibitem [{Note1()}]{Note1}%
  \BibitemOpen
  \bibinfo {note} {See Supplemental Material at for equations of motion of
  active density matrices, further results concerning the impact of model
  parameters on ultrafast exciton dynamics, the mixed quantum/classical
  approach to exciton dynamics, and additional details regarding numerical
  computations of ultrafast pump-probe spectra.}\BibitemShut {Stop}%
\bibitem [{\citenamefont {Hwang}\ \emph {et~al.}(2007)\citenamefont {Hwang},
  \citenamefont {Soci}, \citenamefont {Moses}, \citenamefont {Zhu},
  \citenamefont {Waller}, \citenamefont {Gaudiana}, \citenamefont {Brabec},\
  and\ \citenamefont {Heeger}}]{ADMA:ADMA200602437}%
  \BibitemOpen
  \bibfield  {author} {\bibinfo {author} {\bibfnamefont {I.-W.}\ \bibnamefont
  {Hwang}}, \bibinfo {author} {\bibfnamefont {C.}~\bibnamefont {Soci}},
  \bibinfo {author} {\bibfnamefont {D.}~\bibnamefont {Moses}}, \bibinfo
  {author} {\bibfnamefont {Z.}~\bibnamefont {Zhu}}, \bibinfo {author}
  {\bibfnamefont {D.}~\bibnamefont {Waller}}, \bibinfo {author} {\bibfnamefont
  {R.}~\bibnamefont {Gaudiana}}, \bibinfo {author} {\bibfnamefont {C. J.}\
  \bibnamefont {Brabec}}, \ and\ \bibinfo {author} {\bibfnamefont {A. J.}\
  \bibnamefont {Heeger}},\ }\bibfield  {title} {\enquote {\bibinfo {title}
  {Ultrafast electron transfer and decay dynamics in a small band gap bulk
  heterojunction material},}\ }\href {\doibase 10.1002/adma.200602437}
  {\bibfield  {journal} {\bibinfo  {journal} {Adv. Mater.}\ }\textbf {\bibinfo
  {volume} {19}},\ \bibinfo {pages} {2307--2312} (\bibinfo {year}
  {2007})}\BibitemShut {NoStop}%
\bibitem [{\citenamefont {M{\"u}hlbacher}\ \emph {et~al.}(2006)\citenamefont
  {M{\"u}hlbacher}, \citenamefont {Scharber}, \citenamefont {Morana},
  \citenamefont {Zhu}, \citenamefont {Waller}, \citenamefont {Gaudiana},\ and\
  \citenamefont {Brabec}}]{ADMA:ADMA200600160}%
  \BibitemOpen
  \bibfield  {author} {\bibinfo {author} {\bibfnamefont {D.}~\bibnamefont
  {M{\"u}hlbacher}}, \bibinfo {author} {\bibfnamefont {M.}~\bibnamefont
  {Scharber}}, \bibinfo {author} {\bibfnamefont {M.}~\bibnamefont {Morana}},
  \bibinfo {author} {\bibfnamefont {Z.}~\bibnamefont {Zhu}}, \bibinfo {author}
  {\bibfnamefont {D.}~\bibnamefont {Waller}}, \bibinfo {author} {\bibfnamefont
  {R.}~\bibnamefont {Gaudiana}}, \ and\ \bibinfo {author} {\bibfnamefont
  {C.}~\bibnamefont {Brabec}},\ }\bibfield  {title} {\enquote {\bibinfo {title}
  {High photovoltaic performance of a low-bandgap polymer},}\ }\href {\doibase
  10.1002/adma.200600160} {\bibfield  {journal} {\bibinfo  {journal} {Adv.
  Mater.}\ }\textbf {\bibinfo {volume} {18}},\ \bibinfo {pages} {2884--2889}
  (\bibinfo {year} {2006})}\BibitemShut {NoStop}%
\bibitem [{\citenamefont {Street}\ \emph {et~al.}(2014)\citenamefont {Street},
  \citenamefont {Hawks}, \citenamefont {Khlyabich}, \citenamefont {Li},
  \citenamefont {Schwartz}, \citenamefont {Thompson},\ and\ \citenamefont
  {Yang}}]{JPhysChemC.118.21873}%
  \BibitemOpen
  \bibfield  {author} {\bibinfo {author} {\bibfnamefont {R.~A.}\ \bibnamefont
  {Street}}, \bibinfo {author} {\bibfnamefont {S.~A.}\ \bibnamefont {Hawks}},
  \bibinfo {author} {\bibfnamefont {P.~P.}\ \bibnamefont {Khlyabich}}, \bibinfo
  {author} {\bibfnamefont {G.}~\bibnamefont {Li}}, \bibinfo {author}
  {\bibfnamefont {B.~J.}\ \bibnamefont {Schwartz}}, \bibinfo {author}
  {\bibfnamefont {B.~C.}\ \bibnamefont {Thompson}}, \ and\ \bibinfo {author}
  {\bibfnamefont {Y.}~\bibnamefont {Yang}},\ }\bibfield  {title} {\enquote
  {\bibinfo {title} {Electronic structure and transition energies in
  polymer-fullerene bulk heterojunctions},}\ }\href {\doibase
  10.1021/jp507097h} {\bibfield  {journal} {\bibinfo  {journal} {J. Phys. Chem.
  C}\ }\textbf {\bibinfo {volume} {118}},\ \bibinfo {pages} {21873--21883}
  (\bibinfo {year} {2014})}\BibitemShut {NoStop}%
\bibitem [{\citenamefont {Tamura}\ and\ \citenamefont
  {Tsukada}(2012)}]{PhysRevB.85.054301}%
  \BibitemOpen
  \bibfield  {author} {\bibinfo {author} {\bibfnamefont {H.}~\bibnamefont
  {Tamura}}\ and\ \bibinfo {author} {\bibfnamefont {M.}~\bibnamefont
  {Tsukada}},\ }\bibfield  {title} {\enquote {\bibinfo {title} {Role of
  intermolecular charge delocalization on electron transport in fullerene
  aggregates},}\ }\href {\doibase 10.1103/PhysRevB.85.054301} {\bibfield
  {journal} {\bibinfo  {journal} {Phys. Rev. B}\ }\textbf {\bibinfo {volume}
  {85}},\ \bibinfo {pages} {054301} (\bibinfo {year} {2012})}\BibitemShut
  {NoStop}%
\bibitem [{\citenamefont {\emph{et al.}}(2009)}]{QE-2009}%
  \BibitemOpen
  \bibfield  {author} {\bibinfo {author} {\bibfnamefont {P.~Giannozzi}\
  \bibnamefont {\emph{et al.}}},\ }\bibfield  {title} {\enquote {\bibinfo
  {title} {Quantum espresso: a modular and open-source software project for
  quantum simulations of materials},}\ }\href {\doibase
  10.1088/0953-8984/21/39/395502} {\bibfield  {journal} {\bibinfo  {journal}
  {J. Phys.: Condens. Matter}\ }\textbf {\bibinfo {volume} {21}},\ \bibinfo
  {pages} {395502} (\bibinfo {year} {2009})}\BibitemShut {NoStop}%
\bibitem [{\citenamefont {Kanai}\ and\ \citenamefont
  {Grossman}(2007)}]{NanoLett.7.1967}%
  \BibitemOpen
  \bibfield  {author} {\bibinfo {author} {\bibfnamefont {Y.}~\bibnamefont
  {Kanai}}\ and\ \bibinfo {author} {\bibfnamefont {J.~C.}\ \bibnamefont
  {Grossman}},\ }\bibfield  {title} {\enquote {\bibinfo {title} {Insights on
  interfacial charge transfer across {P}3{H}{T}/fullerene photovoltaic
  heterojunction from ab initio calculations},}\ }\href
  {http://dx.doi.org/10.1021/nl0707095} {\bibfield  {journal} {\bibinfo
  {journal} {Nano Lett.}\ }\textbf {\bibinfo {volume} {7}},\ \bibinfo {pages}
  {1967--1972} (\bibinfo {year} {2007})}\BibitemShut {NoStop}%
\bibitem [{\citenamefont {Lee}\ \emph {et~al.}(2015)\citenamefont {Lee},
  \citenamefont {Arag{\'o}},\ and\ \citenamefont {Troisi}}]{jpcc.119.14989}%
  \BibitemOpen
  \bibfield  {author} {\bibinfo {author} {\bibfnamefont {M.~H.}\ \bibnamefont
  {Lee}}, \bibinfo {author} {\bibfnamefont {J.}~\bibnamefont {Arag{\'o}}}, \
  and\ \bibinfo {author} {\bibfnamefont {A.}~\bibnamefont {Troisi}},\
  }\bibfield  {title} {\enquote {\bibinfo {title} {Charge dynamics in organic
  photovoltaic materials: Interplay between quantum diffusion and quantum
  relaxation},}\ }\href {\doibase 10.1021/acs.jpcc.5b03989} {\bibfield
  {journal} {\bibinfo  {journal} {J. Phys. Chem. C}\ }\textbf {\bibinfo
  {volume} {119}},\ \bibinfo {pages} {14989--14998} (\bibinfo {year}
  {2015})}\BibitemShut {NoStop}%
\bibitem [{\citenamefont {Bittner}\ and\ \citenamefont
  {Ramon}(2007)}]{bittnerramoncont}%
  \BibitemOpen
  \bibfield  {author} {\bibinfo {author} {\bibfnamefont {E.~R.}\ \bibnamefont
  {Bittner}}\ and\ \bibinfo {author} {\bibfnamefont {J.~G.~S.}\ \bibnamefont
  {Ramon}},\ }\bibfield  {title} {\enquote {\bibinfo {title} {Exciton and
  charge-transfer dynamics in polymer semiconductors},}\ }in\ \href@noop {}
  {\emph {\bibinfo {booktitle} {Quantum Dynamics of Complex Molecular
  Systems}}},\ \bibinfo {editor} {edited by\ \bibinfo {editor} {\bibfnamefont
  {D.~A.}\ \bibnamefont {Micha}}\ and\ \bibinfo {editor} {\bibfnamefont
  {I.}~\bibnamefont {Burghardt}}}\ (\bibinfo  {publisher} {Springer-Verlag},\
  \bibinfo {address} {Berlin Heidelberg},\ \bibinfo {year} {2007})\BibitemShut
  {NoStop}%
\bibitem [{\citenamefont {Falke}\ \emph {et~al.}(2014)\citenamefont {Falke},
  \citenamefont {Rozzi}, \citenamefont {Brida}, \citenamefont {Maiuri},
  \citenamefont {Amato}, \citenamefont {Sommer}, \citenamefont {De~Sio},
  \citenamefont {Rubio}, \citenamefont {Cerullo}, \citenamefont {Molinari},\
  and\ \citenamefont {Lienau}}]{Science.344.1001}%
  \BibitemOpen
  \bibfield  {author} {\bibinfo {author} {\bibfnamefont {S.~M.}\ \bibnamefont
  {Falke}}, \bibinfo {author} {\bibfnamefont {C.~A.}\ \bibnamefont {Rozzi}},
  \bibinfo {author} {\bibfnamefont {D.}~\bibnamefont {Brida}}, \bibinfo
  {author} {\bibfnamefont {M.}~\bibnamefont {Maiuri}}, \bibinfo {author}
  {\bibfnamefont {M.}~\bibnamefont {Amato}}, \bibinfo {author} {\bibfnamefont
  {E.}~\bibnamefont {Sommer}}, \bibinfo {author} {\bibfnamefont
  {A.}~\bibnamefont {De~Sio}}, \bibinfo {author} {\bibfnamefont
  {A.}~\bibnamefont {Rubio}}, \bibinfo {author} {\bibfnamefont
  {G.}~\bibnamefont {Cerullo}}, \bibinfo {author} {\bibfnamefont
  {E.}~\bibnamefont {Molinari}}, \ and\ \bibinfo {author} {\bibfnamefont
  {C.}~\bibnamefont {Lienau}},\ }\bibfield  {title} {\enquote {\bibinfo {title}
  {Coherent ultrafast charge transfer in an organic photovoltaic blend},}\
  }\href {\doibase 10.1126/science.1249771} {\bibfield  {journal} {\bibinfo
  {journal} {Science}\ }\textbf {\bibinfo {volume} {344}},\ \bibinfo {pages}
  {1001--1005} (\bibinfo {year} {2014})}\BibitemShut {NoStop}%
\bibitem [{\citenamefont {L{\"u}cke}\ \emph {et~al.}(2016)\citenamefont
  {L{\"u}cke}, \citenamefont {Ortmann}, \citenamefont {Panhans}, \citenamefont
  {Sanna}, \citenamefont {Rauls}, \citenamefont {Gerstmann},\ and\
  \citenamefont {Schmidt}}]{JPhysChemB.120.5572}%
  \BibitemOpen
  \bibfield  {author} {\bibinfo {author} {\bibfnamefont {A.}~\bibnamefont
  {L{\"u}cke}}, \bibinfo {author} {\bibfnamefont {F.}~\bibnamefont {Ortmann}},
  \bibinfo {author} {\bibfnamefont {M.}~\bibnamefont {Panhans}}, \bibinfo
  {author} {\bibfnamefont {S.}~\bibnamefont {Sanna}}, \bibinfo {author}
  {\bibfnamefont {E.}~\bibnamefont {Rauls}}, \bibinfo {author} {\bibfnamefont
  {U.}~\bibnamefont {Gerstmann}}, \ and\ \bibinfo {author} {\bibfnamefont
  {W.~G.}\ \bibnamefont {Schmidt}},\ }\bibfield  {title} {\enquote {\bibinfo
  {title} {Temperature-dependent hole mobility and its limit in crystal-phase
  {P}3{H}{T} calculated from first principles},}\ }\href {\doibase
  10.1021/acs.jpcb.6b03598} {\bibfield  {journal} {\bibinfo  {journal} {J.
  Phys. Chem. B}\ }\textbf {\bibinfo {volume} {120}},\ \bibinfo {pages}
  {5572--5580} (\bibinfo {year} {2016})}\BibitemShut {NoStop}%
\bibitem [{\citenamefont {Vukmirovi{\'c}}\ and\ \citenamefont
  {Wang}(2009)}]{NanoLett.9.3996}%
  \BibitemOpen
  \bibfield  {author} {\bibinfo {author} {\bibfnamefont {N.}~\bibnamefont
  {Vukmirovi{\'c}}}\ and\ \bibinfo {author} {\bibfnamefont {L.-W.}\
  \bibnamefont {Wang}},\ }\bibfield  {title} {\enquote {\bibinfo {title}
  {Charge carrier motion in disordered conjugated polymers: A multiscale ab
  initio study},}\ }\href {http://dx.doi.org/10.1021/nl9021539} {\bibfield
  {journal} {\bibinfo  {journal} {Nano Lett.}\ }\textbf {\bibinfo {volume}
  {9}},\ \bibinfo {pages} {3996--4000} (\bibinfo {year} {2009})}\BibitemShut
  {NoStop}%
\bibitem [{\citenamefont {Cheung}\ and\ \citenamefont
  {Troisi}(2010)}]{jpcc.114.20479}%
  \BibitemOpen
  \bibfield  {author} {\bibinfo {author} {\bibfnamefont {D.~L.}\ \bibnamefont
  {Cheung}}\ and\ \bibinfo {author} {\bibfnamefont {A.}~\bibnamefont
  {Troisi}},\ }\bibfield  {title} {\enquote {\bibinfo {title} {Theoretical
  study of the organic photovoltaic electron acceptor {P}{C}{B}{M}: Morphology,
  electronic structure, and charge localization},}\ }\href
  {http://dx.doi.org/10.1021/jp1049167} {\bibfield  {journal} {\bibinfo
  {journal} {J. Phys. Chem. C}\ }\textbf {\bibinfo {volume} {114}},\ \bibinfo
  {pages} {20479--20488} (\bibinfo {year} {2010})}\BibitemShut {NoStop}%
\bibitem [{\citenamefont {Cheng}\ and\ \citenamefont
  {Silbey}(2008)}]{jcp.128.114713}%
  \BibitemOpen
  \bibfield  {author} {\bibinfo {author} {\bibfnamefont {Y.-C.}\ \bibnamefont
  {Cheng}}\ and\ \bibinfo {author} {\bibfnamefont {R.~J.}\ \bibnamefont
  {Silbey}},\ }\bibfield  {title} {\enquote {\bibinfo {title} {A unified theory
  for charge-carrier transport in organic crystals},}\ }\href {\doibase
  http://dx.doi.org/10.1063/1.2894840} {\bibfield  {journal} {\bibinfo
  {journal} {J. Chem. Phys.}\ }\textbf {\bibinfo {volume} {128}},\ \bibinfo
  {eid} {114713} (\bibinfo {year} {2008})}\BibitemShut {NoStop}%
\bibitem [{\citenamefont {Bera}\ \emph {et~al.}(2015)\citenamefont {Bera},
  \citenamefont {Gheeraert}, \citenamefont {Fratini}, \citenamefont {Ciuchi},\
  and\ \citenamefont {Florens}}]{PhysRevB.91.041107}%
  \BibitemOpen
  \bibfield  {author} {\bibinfo {author} {\bibfnamefont {S.}~\bibnamefont
  {Bera}}, \bibinfo {author} {\bibfnamefont {N.}~\bibnamefont {Gheeraert}},
  \bibinfo {author} {\bibfnamefont {S.}~\bibnamefont {Fratini}}, \bibinfo
  {author} {\bibfnamefont {S.}~\bibnamefont {Ciuchi}}, \ and\ \bibinfo {author}
  {\bibfnamefont {S.}~\bibnamefont {Florens}},\ }\bibfield  {title} {\enquote
  {\bibinfo {title} {Impact of quantized vibrations on the efficiency of
  interfacial charge separation in photovoltaic devices},}\ }\href {\doibase
  10.1103/PhysRevB.91.041107} {\bibfield  {journal} {\bibinfo  {journal} {Phys.
  Rev. B}\ }\textbf {\bibinfo {volume} {91}},\ \bibinfo {pages} {041107}
  (\bibinfo {year} {2015})}\BibitemShut {NoStop}%
\bibitem [{\citenamefont {Tully}(1990)}]{JChemPhys.93.1061}%
  \BibitemOpen
  \bibfield  {author} {\bibinfo {author} {\bibfnamefont {J.~C.}\ \bibnamefont
  {Tully}},\ }\bibfield  {title} {\enquote {\bibinfo {title} {Molecular
  dynamics with electronic transitions},}\ }\href {\doibase
  http://dx.doi.org/10.1063/1.459170} {\bibfield  {journal} {\bibinfo
  {journal} {J. Chem. Phys.}\ }\textbf {\bibinfo {volume} {93}},\ \bibinfo
  {pages} {1061--1071} (\bibinfo {year} {1990})}\BibitemShut {NoStop}%
\bibitem [{\citenamefont {Wang}\ and\ \citenamefont
  {Prezhdo}(2014)}]{JPhysChemLett.5.713}%
  \BibitemOpen
  \bibfield  {author} {\bibinfo {author} {\bibfnamefont {L.}~\bibnamefont
  {Wang}}\ and\ \bibinfo {author} {\bibfnamefont {O.~V.}\ \bibnamefont
  {Prezhdo}},\ }\bibfield  {title} {\enquote {\bibinfo {title} {A simple
  solution to the trivial crossing problem in surface hopping},}\ }\href
  {\doibase 10.1021/jz500025c} {\bibfield  {journal} {\bibinfo  {journal} {J.
  Phys. Chem. Lett.}\ }\textbf {\bibinfo {volume} {5}},\ \bibinfo {pages}
  {713--719} (\bibinfo {year} {2014})}\BibitemShut {NoStop}%
\bibitem [{\citenamefont {Chenel}\ \emph {et~al.}(2014)\citenamefont {Chenel},
  \citenamefont {Mangaud}, \citenamefont {Burghardt}, \citenamefont {Meier},\
  and\ \citenamefont {Desouter-Lecomte}}]{JChemPhys.140.044104}%
  \BibitemOpen
  \bibfield  {author} {\bibinfo {author} {\bibfnamefont {A.}~\bibnamefont
  {Chenel}}, \bibinfo {author} {\bibfnamefont {E.}~\bibnamefont {Mangaud}},
  \bibinfo {author} {\bibfnamefont {I.}~\bibnamefont {Burghardt}}, \bibinfo
  {author} {\bibfnamefont {C.}~\bibnamefont {Meier}}, \ and\ \bibinfo {author}
  {\bibfnamefont {M.}~\bibnamefont {Desouter-Lecomte}},\ }\bibfield  {title}
  {\enquote {\bibinfo {title} {Exciton dissociation at donor-acceptor
  heterojunctions: Dynamics using the collective effective mode representation
  of the spin-boson model},}\ }\href {http://dx.doi.org/10.1063/1.4861853}
  {\bibfield  {journal} {\bibinfo  {journal} {J. Chem. Phys.}\ }\textbf
  {\bibinfo {volume} {140}},\ \bibinfo {pages} {044104} (\bibinfo {year}
  {2014})}\BibitemShut {NoStop}%
\bibitem [{\citenamefont {Pensack}\ and\ \citenamefont
  {Asbury}(2010)}]{JPhysChemLett.1.2255}%
  \BibitemOpen
  \bibfield  {author} {\bibinfo {author} {\bibfnamefont {R.~D.}\ \bibnamefont
  {Pensack}}\ and\ \bibinfo {author} {\bibfnamefont {J.~B.}\ \bibnamefont
  {Asbury}},\ }\bibfield  {title} {\enquote {\bibinfo {title} {Beyond the
  adiabatic limit: Charge photogeneration in organic photovoltaic materials},}\
  }\href {\doibase 10.1021/jz1005225} {\bibfield  {journal} {\bibinfo
  {journal} {J. Phys. Chem. Lett.}\ }\textbf {\bibinfo {volume} {1}},\ \bibinfo
  {pages} {2255--2263} (\bibinfo {year} {2010})}\BibitemShut {NoStop}%
\bibitem [{\citenamefont {Mukamel}(1995)}]{MukamelBook}%
  \BibitemOpen
  \bibfield  {author} {\bibinfo {author} {\bibfnamefont {S.}~\bibnamefont
  {Mukamel}},\ }\href@noop {} {\emph {\bibinfo {title} {Principles of Nonlinear
  Optical Spectroscopy}}}\ (\bibinfo  {publisher} {Oxford University Press},\
  \bibinfo {address} {New York},\ \bibinfo {year} {1995})\BibitemShut {NoStop}%
\bibitem [{\citenamefont {Cabanillas-Gonzalez}\ \emph
  {et~al.}(2011)\citenamefont {Cabanillas-Gonzalez}, \citenamefont {Grancini},\
  and\ \citenamefont {Lanzani}}]{AdvMater.23.5468}%
  \BibitemOpen
  \bibfield  {author} {\bibinfo {author} {\bibfnamefont {J.}~\bibnamefont
  {Cabanillas-Gonzalez}}, \bibinfo {author} {\bibfnamefont {G.}~\bibnamefont
  {Grancini}}, \ and\ \bibinfo {author} {\bibfnamefont {G.}~\bibnamefont
  {Lanzani}},\ }\bibfield  {title} {\enquote {\bibinfo {title} {Pump-probe
  spectroscopy in organic semiconductors: Monitoring fundamental processes of
  relevance in optoelectronics},}\ }\href {\doibase 10.1002/adma.201102015}
  {\bibfield  {journal} {\bibinfo  {journal} {Adv. Mater.}\ }\textbf {\bibinfo
  {volume} {23}},\ \bibinfo {pages} {5468--5485} (\bibinfo {year}
  {2011})}\BibitemShut {NoStop}%
\bibitem [{\citenamefont {Perfetto}\ and\ \citenamefont
  {Stefanucci}(2015)}]{PhysRevA.91.033416}%
  \BibitemOpen
  \bibfield  {author} {\bibinfo {author} {\bibfnamefont {E.}~\bibnamefont
  {Perfetto}}\ and\ \bibinfo {author} {\bibfnamefont {G.}~\bibnamefont
  {Stefanucci}},\ }\bibfield  {title} {\enquote {\bibinfo {title} {Some exact
  properties of the nonequilibrium response function for transient
  photoabsorption},}\ }\href {\doibase 10.1103/PhysRevA.91.033416} {\bibfield
  {journal} {\bibinfo  {journal} {Phys. Rev. A}\ }\textbf {\bibinfo {volume}
  {91}},\ \bibinfo {pages} {033416} (\bibinfo {year} {2015})}\BibitemShut
  {NoStop}%
\bibitem [{\citenamefont {Walkenhorst}\ \emph {et~al.}(2016)\citenamefont
  {Walkenhorst}, \citenamefont {De~Giovannini}, \citenamefont {Castro},\ and\
  \citenamefont {Rubio}}]{EurPhysJB.89.128}%
  \BibitemOpen
  \bibfield  {author} {\bibinfo {author} {\bibfnamefont {J.}~\bibnamefont
  {Walkenhorst}}, \bibinfo {author} {\bibfnamefont {U.}~\bibnamefont
  {De~Giovannini}}, \bibinfo {author} {\bibfnamefont {A.}~\bibnamefont
  {Castro}}, \ and\ \bibinfo {author} {\bibfnamefont {A.}~\bibnamefont
  {Rubio}},\ }\bibfield  {title} {\enquote {\bibinfo {title} {Tailored
  pump-probe transient spectroscopy with time-dependent density-functional
  theory: controlling absorption spectra},}\ }\href {\doibase
  10.1140/epjb/e2016-70064-0} {\bibfield  {journal} {\bibinfo  {journal} {Eur.
  Phys. J. B}\ }\textbf {\bibinfo {volume} {89}},\ \bibinfo {pages} {128}
  (\bibinfo {year} {2016})}\BibitemShut {NoStop}%
\bibitem [{Note2()}]{Note2}%
  \BibitemOpen
  \bibinfo {note} {Coherences between exciton states and the ground state $y_x$
  typically decay on $\lesssim $100-fs time scale after the pump field has
  vanished, while exciton-exciton coherences typically decay on ps time scales
  or longer. Therefore, in our computations, we expect to see the decay of
  coherences between exciton states and the ground state, but not of the
  exciton-exciton coherences, and time scales on which Eq.~\protect \textup
  {\hbox {\mathsurround \z@ \protect \normalfont (\ignorespaces \ref
  {Eq:IV_sab}\unskip \@@italiccorr )}} is valid are in principle at least ps or
  longer.}\BibitemShut {Stop}%
\bibitem [{\citenamefont {Valiev}\ \emph {et~al.}(2010)\citenamefont {Valiev},
  \citenamefont {Bylaska}, \citenamefont {Govind}, \citenamefont {Kowalski},
  \citenamefont {Straatsma}, \citenamefont {Dam}, \citenamefont {Wang},
  \citenamefont {Nieplocha}, \citenamefont {Apra}, \citenamefont {Windus},\
  and\ \citenamefont {de~Jong}}]{cpc181-1477}%
  \BibitemOpen
  \bibfield  {author} {\bibinfo {author} {\bibfnamefont {M.}~\bibnamefont
  {Valiev}}, \bibinfo {author} {\bibfnamefont {E.J.}\ \bibnamefont {Bylaska}},
  \bibinfo {author} {\bibfnamefont {N.}~\bibnamefont {Govind}}, \bibinfo
  {author} {\bibfnamefont {K.}~\bibnamefont {Kowalski}}, \bibinfo {author}
  {\bibfnamefont {T.P.}\ \bibnamefont {Straatsma}}, \bibinfo {author}
  {\bibfnamefont {H.J.J.~Van}\ \bibnamefont {Dam}}, \bibinfo {author}
  {\bibfnamefont {D.}~\bibnamefont {Wang}}, \bibinfo {author} {\bibfnamefont
  {J.}~\bibnamefont {Nieplocha}}, \bibinfo {author} {\bibfnamefont
  {E.}~\bibnamefont {Apra}}, \bibinfo {author} {\bibfnamefont {T.L.}\
  \bibnamefont {Windus}}, \ and\ \bibinfo {author} {\bibfnamefont {W.A.}\
  \bibnamefont {de~Jong}},\ }\bibfield  {title} {\enquote {\bibinfo {title}
  {{N}{W}{C}hem: A comprehensive and scalable open-source solution for large
  scale molecular simulations},}\ }\href {\doibase 10.1016/j.cpc.2010.04.018}
  {\bibfield  {journal} {\bibinfo  {journal} {Comput. Phys. Commun.}\ }\textbf
  {\bibinfo {volume} {181}},\ \bibinfo {pages} {1477} (\bibinfo {year}
  {2010})}\BibitemShut {NoStop}%
\bibitem [{\citenamefont {Kocherzhenko}\ \emph {et~al.}(2015)\citenamefont
  {Kocherzhenko}, \citenamefont {Lee}, \citenamefont {Forsuelo},\ and\
  \citenamefont {Whaley}}]{jpcc.119.7590}%
  \BibitemOpen
  \bibfield  {author} {\bibinfo {author} {\bibfnamefont {A.~A.}\ \bibnamefont
  {Kocherzhenko}}, \bibinfo {author} {\bibfnamefont {D.}~\bibnamefont {Lee}},
  \bibinfo {author} {\bibfnamefont {M.~A.}\ \bibnamefont {Forsuelo}}, \ and\
  \bibinfo {author} {\bibfnamefont {K.~B.}\ \bibnamefont {Whaley}},\ }\bibfield
   {title} {\enquote {\bibinfo {title} {Coherent and incoherent contributions
  to charge separation in multichromophore systems},}\ }\href {\doibase
  10.1021/jp5127859} {\bibfield  {journal} {\bibinfo  {journal} {J. Phys. Chem.
  C}\ }\textbf {\bibinfo {volume} {119}},\ \bibinfo {pages} {7590--7603}
  (\bibinfo {year} {2015})}\BibitemShut {NoStop}%
\bibitem [{\citenamefont {Koster}\ \emph {et~al.}(2006)\citenamefont {Koster},
  \citenamefont {Mihailetchi},\ and\ \citenamefont
  {Blom}}]{ApplPhysLett.88.093511}%
  \BibitemOpen
  \bibfield  {author} {\bibinfo {author} {\bibfnamefont {L.~J.~A.}\
  \bibnamefont {Koster}}, \bibinfo {author} {\bibfnamefont {V.~D.}\
  \bibnamefont {Mihailetchi}}, \ and\ \bibinfo {author} {\bibfnamefont
  {P.~W.~M.}\ \bibnamefont {Blom}},\ }\bibfield  {title} {\enquote {\bibinfo
  {title} {Ultimate efficiency of polymer/fullerene bulk heterojunction solar
  cells},}\ }\href {http://dx.doi.org/10.1063/1.2181635} {\bibfield  {journal}
  {\bibinfo  {journal} {Appl. Phys. Lett.}\ }\textbf {\bibinfo {volume} {88}},\
  \bibinfo {pages} {093511} (\bibinfo {year} {2006})}\BibitemShut {NoStop}%
\end{thebibliography}%

\end{document}